\documentclass[preprint,aps,pre,showpacs,epsfig] {revtex4-1}
\usepackage{hyperref}
\usepackage{pdfsync}
\usepackage{graphicx}   
\usepackage{epstopdf}
\usepackage{amsmath}

\begin{document}

\title{A New Twist on the Electroclinic Critical Point: Type I and Type II Smectic $C^*$ Systems}

\author{Josh Ziegler}
\altaffiliation[Current address: ]{Department of Physics, University of Oregon, Eugene, OR 97403, USA}
\author{Sean Echols}
\author{Matthew J. Moelter}
\author{Karl Saunders}
\email{ksaunder@calpoly.edu}
\affiliation{Department of Physics \\ California Polytechnic State University \\ San Luis Obispo, CA 93407, USA}

\date{\today}

\begin{abstract}

We conduct an in-depth analysis of the electroclinic effect in ferroelectric liquid crystal systems that have a first order Smectic-$A^*$--Smectic-$C^*$ (Sm-$A^*$--Sm-$C^*$) transition, and show that such systems can be either Type I or Type II. In temperature--field parameter space Type I systems exhibit a macroscopically achiral (in which the Sm-$C^*$ helical superstructure is expelled) low-tilt (LT) Sm-$C$--high-tilt (HT) Sm-$C$ critical point, which terminates a LT Sm-$C$--HT Sm-$C$ first order boundary. This boundary extends to an achiral-chiral triple point at which the achiral LT Sm-$C$ and HT Sm-$C$ phases coexist along with the chiral Sm-$C^*$ phase. In Type II systems the critical point, triple point, and first order boundary are replaced by a Sm-$C^*$ region, sandwiched between LT and HT achiral Sm-$C$ phases, at low and high fields respectively. Correspondingly, as field is ramped up, the Type II system will display a reentrant Sm-$C$--Sm-$C^*$-Sm-$C$ phase sequence. Moreover, discontinuity in the tilt of the optical axis at of the two each phase transitions means the Type II system is tristable. This is in contrast to the bistable nature of the LT Sm-$C$--HT Sm-$C$ transition in Type I systems. Whether the system is Type I or Type II is determined by the ratio of two length scales, one of which is the zero-field Sm-$C^*$ helical pitch. The other length scale depends on the size of the discontinuity (and thus the latent heat) at the zero-field first order Sm-$A^*$--Sm-$C^*$ transition. We propose ways in which a system could be experimentally tuned, e.g., by varying enantiomeric excess, between Type I and Type II behavior. We also show that this Type I vs Type II behavior is the Ising universality class analog of Type I vs Type II behavior in XY universality class systems. Specifically, the LT and HT achiral Sm-$C$ phases are analogous to normal and superconducting phases, while the 1D periodic Sm-$C^*$ superstructure is analogous to the 2D periodic Abrikosov flux lattice. Lastly, we make a complete mapping of the phase boundaries in all regions of temperature--field-enantiomeric excess parameter space (not just near the critical point) and show that a variety of interesting features are possible, including a multicritical point, tricritical points and a doubly reentrant Sm-$C$--Sm-$C^*$-Sm-$C$--Sm-$C^*$ phase sequence. 

\end{abstract}


\maketitle


\section{Introduction: Chiral Smectic $C$ Phases and the Electroclinic Critical Point}
\label{sec:Introduction}

In the 1970s Meyer\cite{Meyer} used an elegant symmetry argument to predict that the application of an electric field to a chiral Smectic-$A^*$ (Sm-$A^*$) phase would induce a transition to the Smectic-$C^*$ (Sm-$C^*$) phase, along with an associated tilt of the optical axis. This electroclinic effect was subsequently confirmed experimentally\cite{Garoff and Meyer}. This in turn led to the development of electro-optic devices using ferroelectric (chiral) liquid crystals and also to the synthesis of many new ferroelectric liquid crystals possessing Sm-$A^*$ and Sm-$C^*$ phases. These ferroelectric liquid crystals, and their behavior in an electric field, have been the subject of much experimental and theoretical study in the decades since the discovery of the 
electroclinic effect\cite{Ferroelectric Review}.

\begin{figure}[ht]
\includegraphics[scale=0.5]{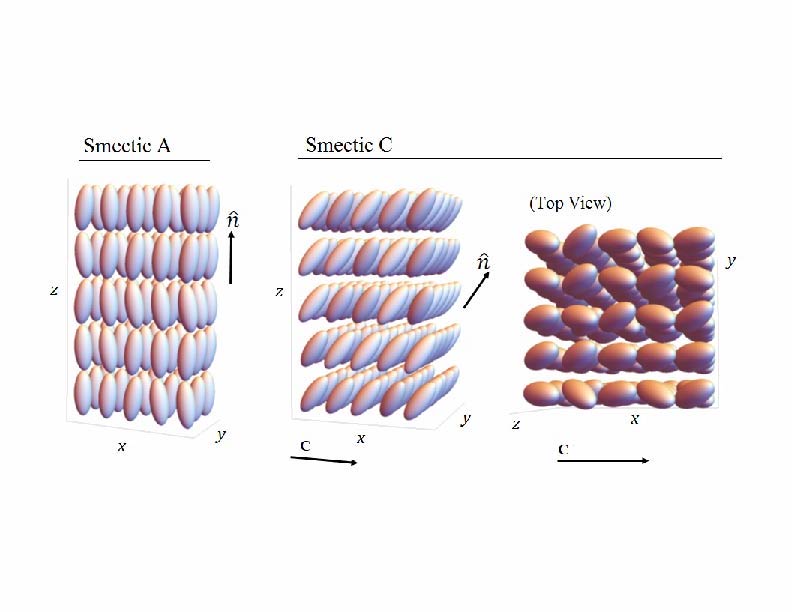}
\caption{A schematic representation of the Sm-$A$ and Sm-$C$ phases.
\label{Smectic Schematic}}
\end{figure}

It is worth briefly reviewing the basics of the smectic phases. Smectics have a modulated density along one direction ($\bf \hat z$), as shown schematically in Fig.~\ref{Smectic Schematic}. The elongated molecules tend to align their long axes along a common direction ($\bf \hat n$) known as the director or optical axis. In the Sm-$A$ phase, ${\bf \hat n} = {\bf \hat z}$, while in the lower temperature Sm-$C$ phase, ${\bf \hat n}$ lies at an angle relative to ${\bf \hat z}$. The order parameter of the Sm-$C$ phase is $\bf c$, the projection of ${\bf \hat n}$ onto the $xy$ layering plane. For nonchiral smectics, the transition from the Sm-$A$ to the Sm-$C$ phase is typically induced by lowering temperature, but can also be induced by compressing the smectic layers or by varying concentration. For a chiral Sm-$A$ phase, the transition can also be induced via the electroclinic effect, i.e, by applying an electric field $\bf E$ perpendicular to $\bf \hat z$. Application of a field to a system already in the Sm-$C^*$ phase will increase the degree of tilt. It is important to note that the direction of the induced tilt ($\bf c$) is determined by the direction of $\bf E$. For the schematic shown in Fig.~\ref{Smectic Schematic}, ${\bf E} = E{\bf \hat y}$ (into the page) induces ${\bf c} = c{\bf \hat x}$ (to the right). Switching the field, i.e.,  ${\bf E} = -E{\bf \hat y}$ (out of the page) would switch the direction of the induced tilt, i.e., ${\bf c} = -c{\bf \hat x}$ (to the left). This switching behavior is a key feature in the operation of surface stabilized ferroelectric LCDs\cite{Clark and Lagerwall}.

\begin{figure}[ht]
\includegraphics[scale=0.5]{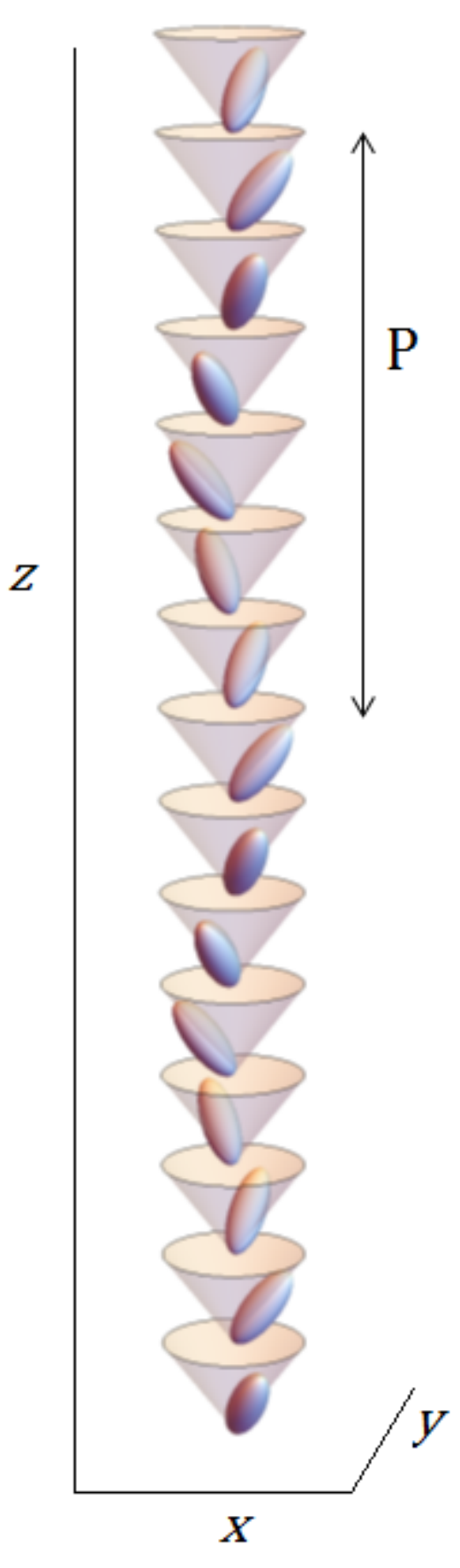}
\caption{A schematic representation the helical Sm-$C^*$ phase, in which the average molecular direction precesses from layer to layer. This modulation along $z$ has period $P$.
\label{Helical Sm-C* Phase}}
\end{figure}

The simple situation described above is complicated by the fact that in the Sm-$C^*$ phase, the director $\bf \hat n$ is not uniform. Instead, it precesses around $\bf \hat z$ with a periodicity that is larger than the layer spacing. In the absence of a field, this precession is helical, as shown schematically in Fig.~\ref{Helical Sm-C* Phase}. This Sm-$C^*$ chiral superstructure is a macroscopic manifestation of the microscopic chirality of the constituent molecules\cite{Sm-A* no macro chirality footnote}. As discussed above, the electroclinic effect both increases the magnitude of the tilt {\it and} selects a direction for the tilt. This means that the field tends to expel the Sm-$C^*$ chiral superstructure, and for a sufficiently large field, there is a transition from the Sm-$C^*$ phase to the Sm-$C$ phase in which the director $\bf \hat n$ is uniform. Notationally, we use Sm-$C^*$ for the phase in which $\bf \hat n$ is modulated and precesses around $\bf \hat z$, while we use Sm-$C$ for the phase in which $\bf \hat n$ is uniform and the macroscopic chiral superstructure has been expelled. Of course, the Sm-$C$ phase is still microscopically chiral and is thus responsive to the electroclinic effect. We will often refer to this Sm-$C^*$--Sm-$C$ transition as a modulated--uniform transition. It is important to remember that modulated--uniform refers to the director $\bf \hat n$ and not to the density. In both the Sm-$C^*$ and Sm-$C$ phases the density is modulated. 

\begin{figure}[ht]
\includegraphics[scale=0.5]{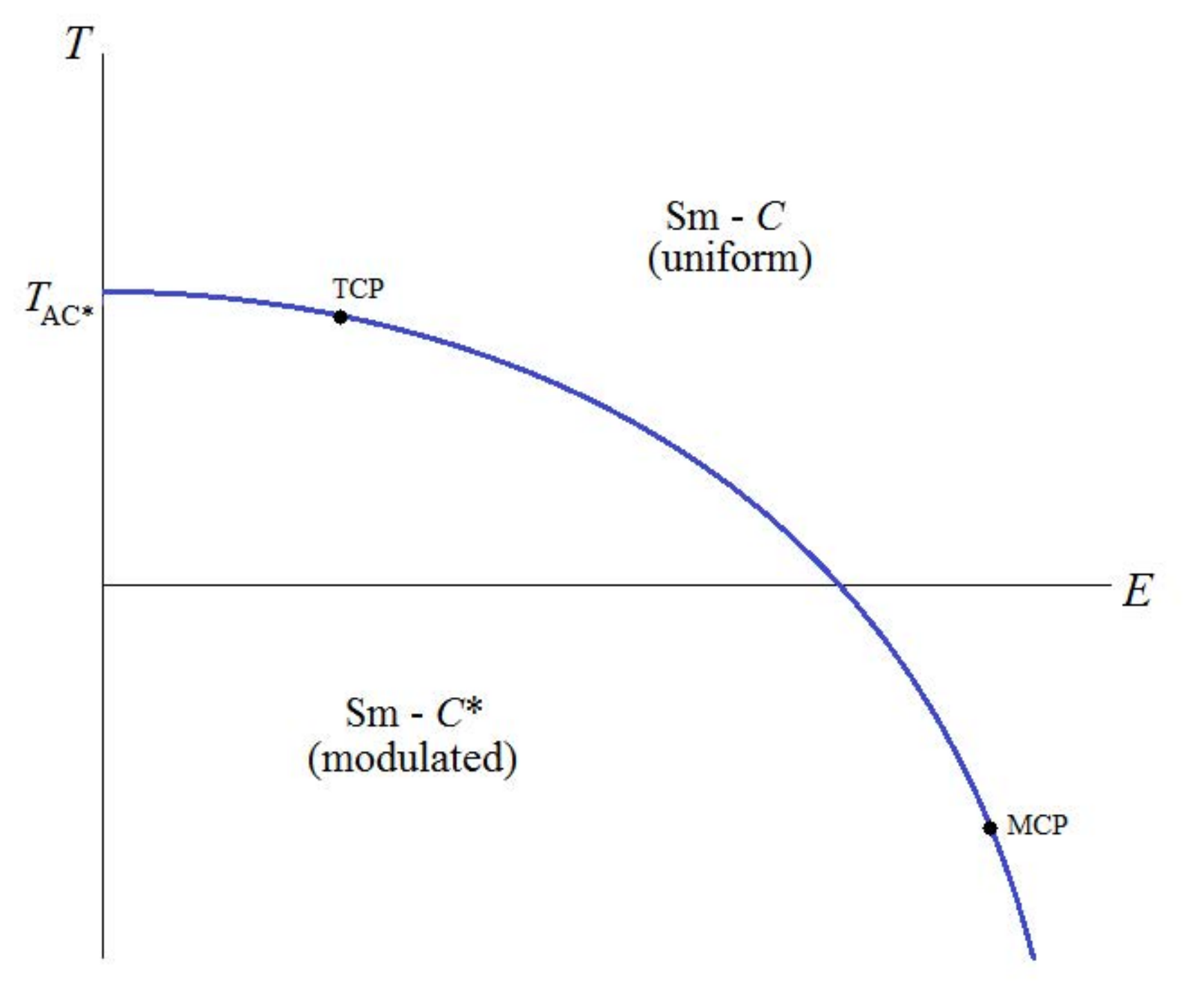}
\caption{The temperature ($T$)-- electric field ($E$) phase diagram for systems with a {\em continuous} zero-field Sm-$A$--Sm-$C^*$ transition at $T_{AC^*}$, as found in \cite{Schaub and Mukamel}. The phase boundary at non-zero field separates the modulated Sm-$C^*$ and uniform Sm-$C$ phases. For temperatures above the tricritical point (TCP) the Sm-$C^*$--Sm-$C$ transition is continuous. Below the TCP it is first order. For temperatures below the multicritical point (MCP) the transition from the Sm-$C^*$ phase to the Sm-$C$ occurs via unwinding of the modulated structure, i.e., a divergence of the modulation period. 
\label{T-E Phase Diagram 2nd order}}
\end{figure}

The critical field $E_c(T)$ for the transition between the Sm-$C^*$ and Sm-$C$ phases depends on temperature $T$. In the simplest Landau model, proposed by Schaub and Mukamel \cite{Schaub and Mukamel}, the critical field $E_c(T)$ grows monotonically with decreasing temperature, as shown in Fig.~\ref{T-E Phase Diagram 2nd order}. In other words, the deeper into the Sm-$C^*$ phase, the larger the field required to expel the modulated superstructure. The $T$--$E$ phase diagram of Fig.~\ref{T-E Phase Diagram 2nd order} also indicates that the application of a field above the Sm-$A$--Sm-$C^*$ transition temperature $T_{AC}$ results in the uniform, Sm-$C$, phase. The same model \cite{Schaub and Mukamel} also predicts the tricritical point and multicritical points shown in Fig.~\ref{T-E Phase Diagram 2nd order}. Subsequent Landau models \cite{Benguigui, Kutnjak1} included extra terms that result in a non-monotonic $E_c(T)$. These theoretical models showed mixed agreement with the experimental investigations of two compounds DOBA-1-MPC \cite{Levstik} and CE8 \cite{Ghoddoussi, Kutnjak2}.

It is important to note that all of the above theoretical and experimental work applied only to systems that have a {\it continuous} zero-field Sm-$A$--Sm-$C^*$ transition, i.e., a continuous growth of the order parameter magnitude $|{\bf c}|$ upon entry to the Sm-$C^*$ phase. In this article we present $T$--$E$ phase diagrams for systems with a {\em first order} Sm-$A$--Sm-$C^*$ transition, in which $|{\bf c}|$ jumps to a non-zero value upon entry to the Sm-$C^*$ phase, with an associated non-zero latent heat. We will see that the phase diagrams for such systems are considerably richer, including the possibility of reentrance, as well as distinct Type I and Type II behaviors. 

\begin{figure}[ht]
\includegraphics[scale=0.5]{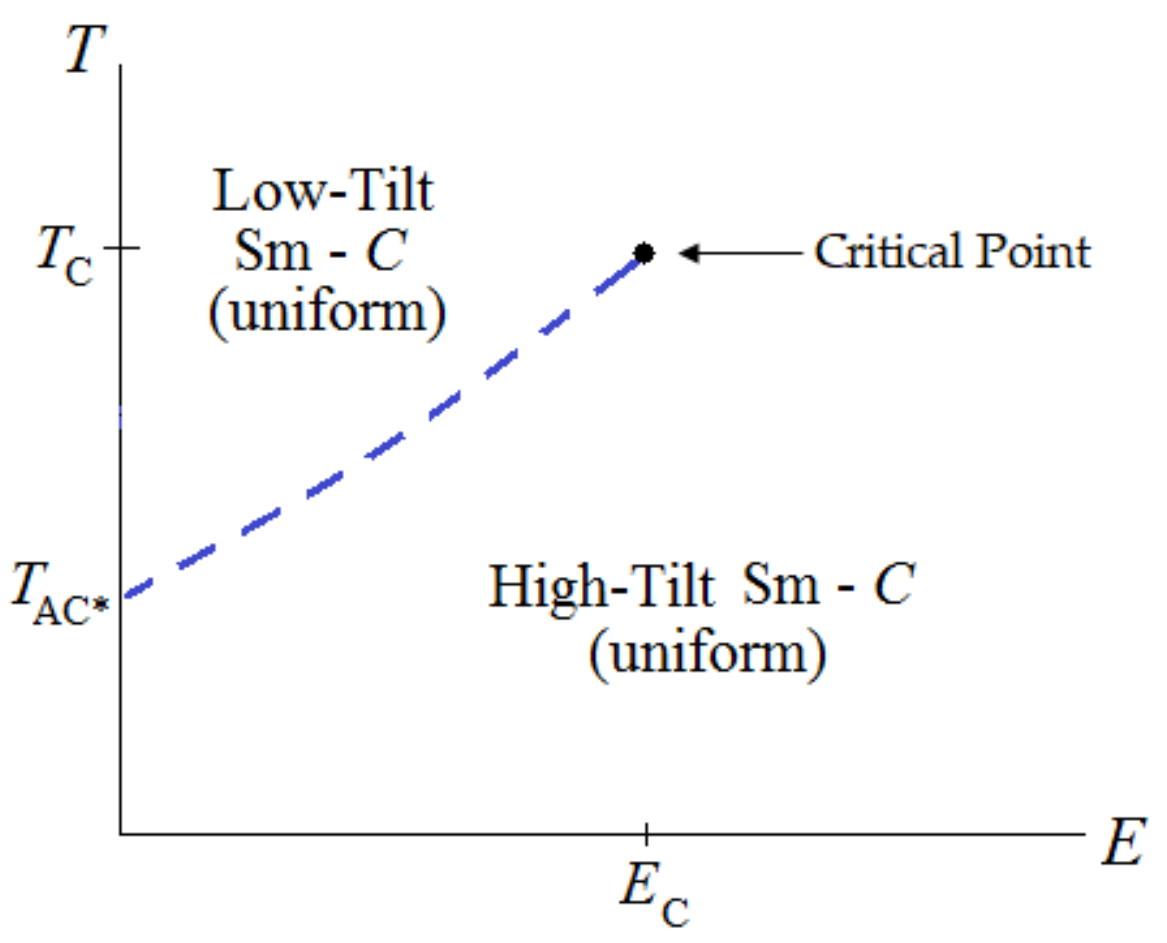}
\caption{The first order phase boundary separating uniform low-tilt from uniform high-tilt Sm-$C$ phases. The phase boundary terminates at an Ising-like critical point, as found in \cite{Bahr and Heppke}. The phase diagram assumes zero modulation.
\label{BECE phase diagram}}
\end{figure}

One motivation for studying systems with a first order Sm-$A$--Sm-$C^*$ transition is that the first order nature of the transition leads to a more dramatic electroclinic effect\cite{Bahr and Heppke}, as shown in Fig.~\ref{Summary of tilt along paths}(a). For materials with a continuous Sm-$A$--Sm-$C^*$ transition, the induced tilt grows continuously with increasing field. For materials with a first order Sm-$A$--Sm-$C^*$ transition, there is a temperature window within which the induced tilt jumps discontinuously with increasing field. This high-tilt--low-tilt phase boundary is shown in Fig.~\ref{BECE phase diagram}. If one ignores the possibility of helical modulation, then the high-tilt--low-tilt phase boundary extends from the (zero field) Sm-$A$--Sm-$C^*$ transition temperature and terminates at a critical point, where the size of the high-tilt--low-tilt discontinuity shrinks to zero. This critical point is basically Ising-like and analogous to the liquid--gas critical point\cite{Prost footnote}.

The theoretical analysis\cite{Bahr and Heppke} of the critical point assumed a Sm-$C^*$ phase without any modulation of the director which, using our notation, means that the analysis considered the electroclinic effect in only the uniform, Sm-$C$ phase. This was justified by the fact that the critical point was at a field larger than the field required to expel the modulated superstructure. However, there is no reason, a priori, to assume that this is true in all systems.

\section{Summary of Results}
\label{sec:SummaryResults}

\begin{figure}[ht]
\includegraphics[scale=0.5]{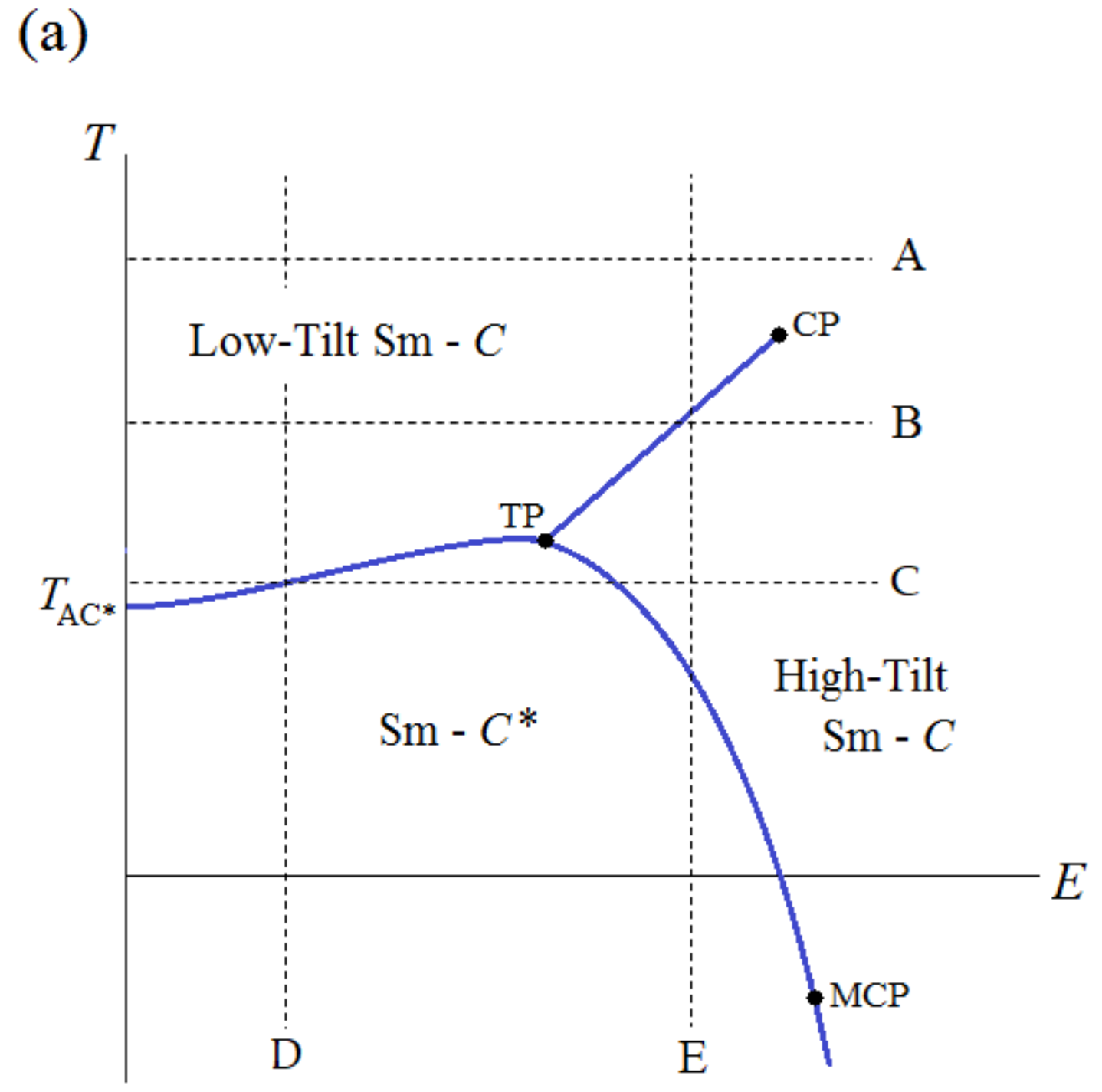}
\includegraphics[scale=0.5]{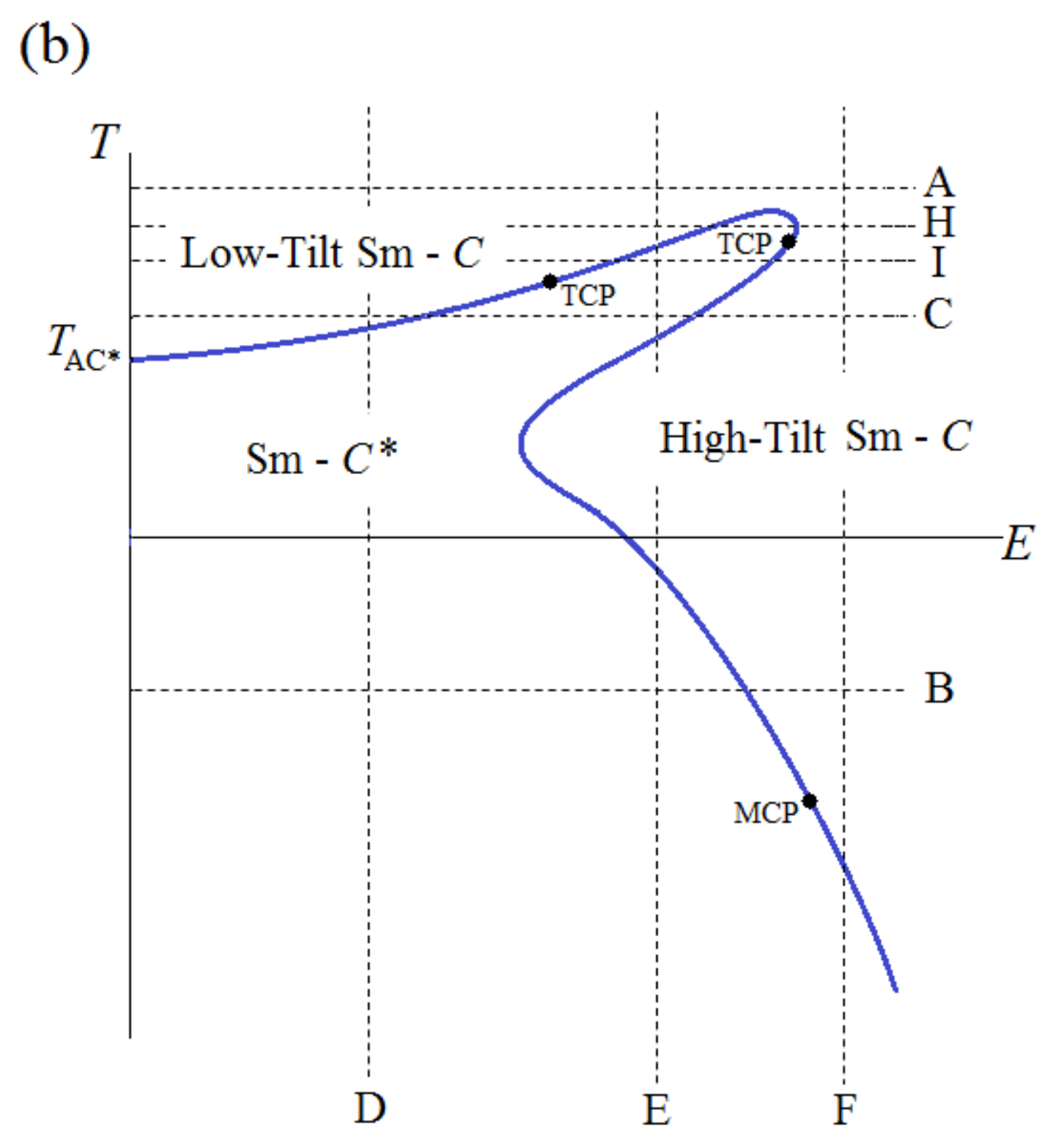}
\caption{$E$-$T$ phase diagrams for (a) Type I systems and (b) Type II systems. The smectic $A$ phase exists for $E=0$ and $T>T_{AC^*}$. In Fig. (a) $CP$ denotes the critical point, $TP$ denotes the triple point and $MCP$ denotes the multicritical point. In Fig. (b) $TCP$ denote the tricritical points. Each figure shows various fixed $T$ and fixed $E$ paths which correspond to different phase sequences. These paths are discussed in the text.
\label{Summary of E-T phase diagrams}}
\end{figure}

In mapping out the $T$--$E$ phase diagrams for systems with a first order Sm-$A$--Sm-$C^*$ transition we are also able to carry out a more complete analysis of the electroclinic effect in such systems, i.e., an analysis that (unlike \cite{Bahr and Heppke}) does not assume a uniform Sm-$C$ phase. We find that for some systems, which we categorize as Type I, the critical point does indeed exist within the uniform (Sm-$C$) region, as asserted in\cite{Bahr and Heppke}. However, we show that for other systems, which we categorize as Type II, the critical point no longer exists, and that the high-tilt--low-tilt line is replaced by a modulated region. Phase diagrams for these two classes of systems are shown in Fig.~\ref{Summary of E-T phase diagrams}.
%

%
\begin{figure}[ht]
\includegraphics[scale=0.45]{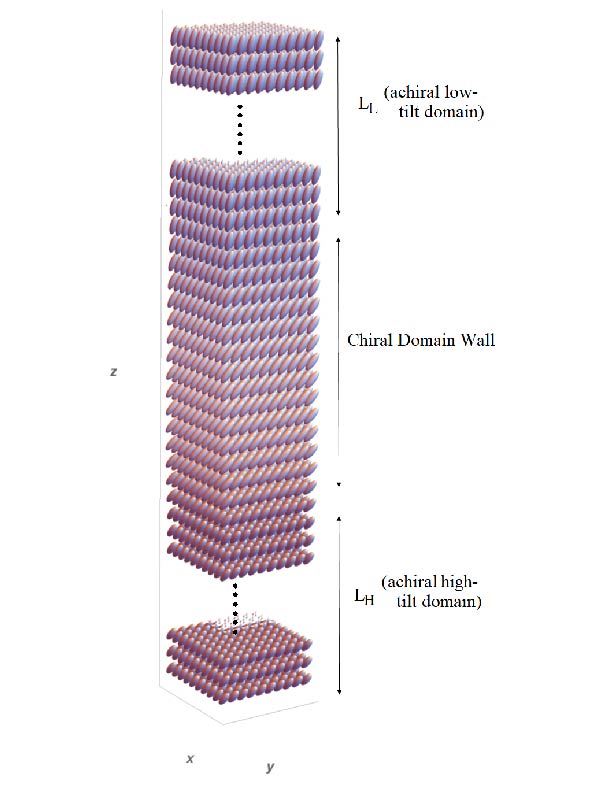}
\caption{Schematic illustration of the modulated Sm-$C^*$ structure in a Type I system close to the triple point. The structure consists of long domains ($L_L\approx P/2$ and $L_H\approx P/2$ of achiral low tilt and high tilt, with the domains separated by short $\ll P$ chiral domain walls in which the system rapidly twists from low to high tilt (or vice versa).
\label{Domain Walls}}
\end{figure}
Whether a system with a first order Sm-$A$--Sm-$C^*$ transition is Type I or Type II depends on a ratio of length scales, which we denote $m$. One of these is the helical pitch at the zero field Sm-$A$--Sm-$C^*$ transition. The other length scale is the correlation length (at the critical point) for fluctuations of the tilt director away from the direction imposed by the electric field. Systems with long helical pitch (corresponding to small chirality) will be Type I, while those with short helical pitch (corresponding to large chirality) will be Type II. Thus, a system could be tuned via varying enantiomeric excess to be Type I or Type II. Of course, the length scale ratio also depends on other system parameters, including, as we shall see, the size of the tilt discontinuity at the first order transition. We will show that systems with weakly first order transitions can be Type II, even with relatively low chirality. In Section \ref{sec:EstimatingLengthScaleRatio} will present a more detailed discussion of the length scale ratio $m$, and also how Type I vs Type II behavior may be accessed experimentally. 

The Type I vs Type II behavior in Sm-$C^*$ phases is reminiscent of Type I vs Type II behavior in superconductors\cite{Abrikosov}. In Type I superconductors there are two phases, the high-$T$ normal phase and low-$T$ superconducting phase. In Type II superconductors there is an intermediate, modulated phase, the Abrikosov flux lattice. Type I vs Type II behavior is determined by the Ginzberg parameter, a ratio of the London penetration depth to the superconducting correlation length. There is also Type I vs Type II behavior in another type of chiral liquid crystal, near the cholesteric ($N^*$)--smectic-$A^*$ transition\cite{RennLubensky}. For Type I systems there is a high-$T$ ($N^*$)--low-$T$ (Sm-$A^*$) transition. For Type II systems there is an intermediate, modulated phase known as the Twist Grain Boundary (TGB) phase, a set of Sm-$A^*$ domains separated by a modulated, periodic array of twist grain boundaries. In this case the Ginzburg parameter is the ratio of the twist penetration depth to the smectic correlation length. For the Type I vs Type II behavior in Sm-$C^*$ systems that we present here, the high-$T$ phase is the low tilt Sm-$C$ while the low-$T$ phase is the high tilt Sm-$C$. Thus, the intermediate modulated Sm-$C^*$ phase is analogous to the Abrikosov flux lattice or the TGB phase. 

It should be pointed out that our Type I vs Type II analogy is only appropriate for temperatures near the critical point. For sufficiently low temperatures the phase will always be modulated. For a Type I system the phase boundary between low tilt and high tilt uniform states intersects the boundary for the modulated phase at a triple point as shown in Fig.~\ref{Summary of E-T phase diagrams}(a). At this uniform-modulated triple point the uniform low-tilt phase, the uniform high-tilt phase, and the modulated phase are all energetically equivalent, and thus the system is equally likely to be found in any of the three phases. 

This uniform-modulated triple point in is analyzed in Section \ref{sec:MappingModulatedPhaseBndyTypeIMUTriplePoint} and we show that it can actually thought of as a Type I--Type II triple point, with the system being Type I on the high $T$ side of the triple point and Type II on the low $T$ side. On the low-$T$ side of the triple point the modulated state is bounded by two first order phase boundaries to the low and high tilt uniform states. Close to the triple point the modulated structure consists of long domains ($L_L\approx P/2$ and $L_H\approx P/2$, where $P$ is the period) of achiral low tilt and high tilt, with the achiral domains separated by short ($\ll P$) chiral domain walls in which the system rapidly twists from low to high tilt (or vice versa), as shown schematically in Fig. \ref{Domain Walls}. At each of the first order boundaries the transition to the modulated phase occurs via the nucleation of a periodic array of chiral domain walls. For example, at the transition from the uniform high-tilt phase to the modulated phase, there is a nucleation of a periodic array of low-tilt domains. Near this high-tilt boundary the modulated phase can be though of as a low $T$ (high tilt) phase riddled with a periodic array of high $T$ (low tilt) defects. This is analogous to the modulated Abrikosov flux lattice in Type II superconductors which can be thought of as a low $T$ (superconducting) phase riddled with a periodic array of high $T$ (normal) defects. As temperature is raised above the high-tilt -- modulated phase boundary, the density of these low-tilt domains grows, and eventually the system transitions to the low-tilt phase. Similarly the transition from the uniform low-tilt phase to the modulated phase occurs via the nucleation of an array of periodic array of high-tilt domains. 

We note that the periodic arrangement of defects in the Abrikosov flux lattice (and the TGB phase) is two dimensional, whereas the periodic defects (chiral domain walls) in the modulated Sm-$C^*$ phase is one dimensional. This difference in the defect structure of the two modulated phases can be ascribed to a difference in symmetry between the two systems. The order parameter that distinguishes the normal and superconducting phases (or the cholesteric and smectic $A$ phases) systems has two components with a transition that can be loosely\cite{Non local U(1) footnote} categorized in the $XY$ universality class. The order parameter distinguishing the low and high tilt phases has one component, and the transition is thus in the Ising universality class. The number of components in the order parameter distinguishing the low $T$ and high $T$ phases thus matches the dimensionality of the defect periodicity in the modulated phase between the low $T$ and high $T$ phases.

It is interesting to consider how the phase diagram for a Sm-$C^*$ Type II system morphs into that for a Type I system, as the length scale ratio $m$ is lowered through a critical value $m_c$, e.g., by reducing the chirality (enantiomeric excess). Specifically, we consider how the Type I critical point and triple point come into existence at $m_c$. Looking at Fig.~\ref{Summary of E-T phase diagrams}(b) we see that the Type II phase diagram has two tricritical points (at temperatures $T_{TC1}$ and $T_{TC2}$) on each side of the ``nose." Above each tricritical point, i.e., for  $T>T_{TC1/2}$, the transition between the uniform phase and modulated phase is continuous, while below each tricritical point ($T<T_{TC1/2}$) it is first order. As $m$ is lowered towards $m_c$ the nose narrows, the tricritical points approach one another and the continuous phase boundary between them shrinks. Eventually at $m=m_c$ the continuous phase boundary vanishes, with the two tricritical points merging to form the critical point. As $m$ is lowered below $m_c$, the triple point emerges from the critical point, with the two points connected by a first order boundary between the uniform low and high tilt phases. As $m$ is lowered further the length of this first order boundary grows. Section \ref{sec:CrossoverBetweenTypeITypeII} provides a more detailed discussion as well as a 3D phase diagram, Fig.~\ref{3DPhaseDiagram}, in temperature-field-chirality parameter space.

Another notable result of our model is the possibility of uniform-modulated reentrance within Type I or Type II systems, for which there is no analogous behavior in systems with normal-superconducting or $N^*$--Sm-$A^*$ transitions\cite{Modulated Magnetic Systems Footnote}. Unlike Fig.~\ref{T-E Phase Diagram 2nd order} the phase boundary separating the modulated Sm-$C^*$ and uniform Sm-$C$ phases is non-monotonic.  This makes reentrance possible. Several fixed $T$ paths are shown in Fig.~\ref{Summary of E-T phase diagrams} (a) and (b) for Type I and Type II systems respectively.  Each path corresponds to the field being ramped up from zero, i.e., starting in the Sm-$A$ phase (which only exists at zero field). In Fig.~\ref{Summary of E-T phase diagrams} (a) Paths A and B correspond to the non-reentrant phase sequences Sm-$A$--Sm-$C$ and Sm-$A$--Sm-$C$--Sm-$C$ respectively, while Path C will exhibit the reentrant phase sequence Sm-$A$--Sm-$C$--Sm-$C^*$--Sm-$C$. In Fig.~\ref{Summary of E-T phase diagrams}(b) along paths A and B the system will exhibit the non-reentrant phase sequences Sm-$A$--Sm-$C$ and Sm-$C^*$--Sm-$C$ respectively. Path C will exhibit the reentrant phase sequences Sm-$A$--Sm-$C$--Sm-$C^*$--Sm-$C$. 

Also shown in Fig.~\ref{Summary of E-T phase diagrams}(a) and (b) are several fixed $E$ paths. Each of these corresponds to the temperature being reduced from a starting temperature above the Sm-$A$--Sm-$C^*$ transition. For both Type I and II systems, reducing the temperature at zero field will take the system through the first order Sm-$A$--Sm-$C^*$ transition. In Fig.~\ref{Summary of E-T phase diagrams}(a) Paths D (Sm-$C$--Sm-$C^*$) and E (Sm-$C$-Sm-$C$--Sm-$C^*$) are non-reentrant. In Fig.~\ref{Summary of E-T phase diagrams}(b) Paths D and F take the system from the uniform Sm-$C$ phase into modulated Sm-$C^*$ phase without any reentrance. The intermediate Path E also takes the system from the uniform Sm-$C$ phase into the modulated Sm-$C^*$ phase, but through the {\it doubly reentrant} phase sequence Sm-$C$--Sm-$C^*$--Sm-$C$--Sm-$C^*$.

\begin{figure}[ht]
\includegraphics[scale=0.5]{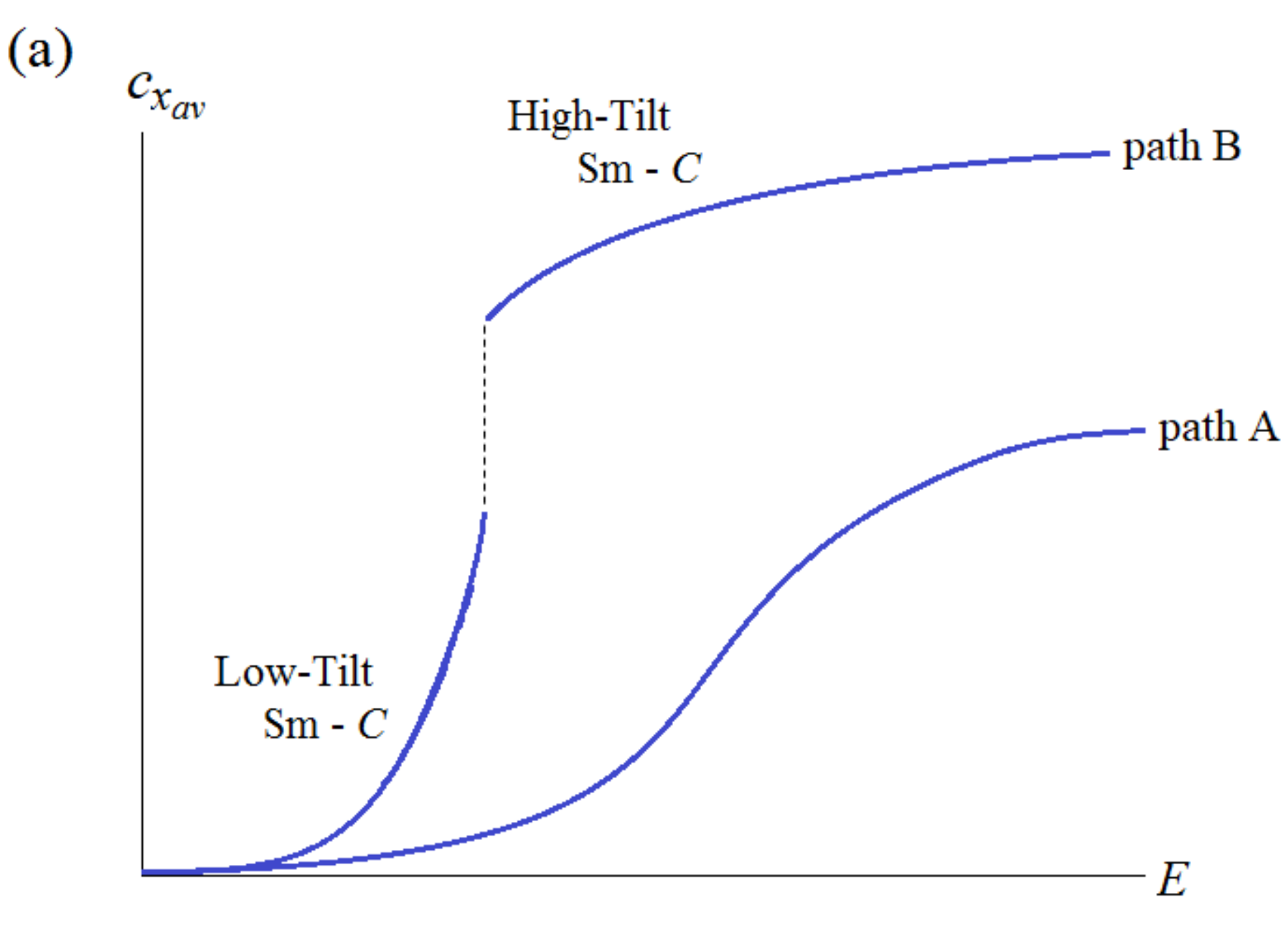}
\includegraphics[scale=0.5]{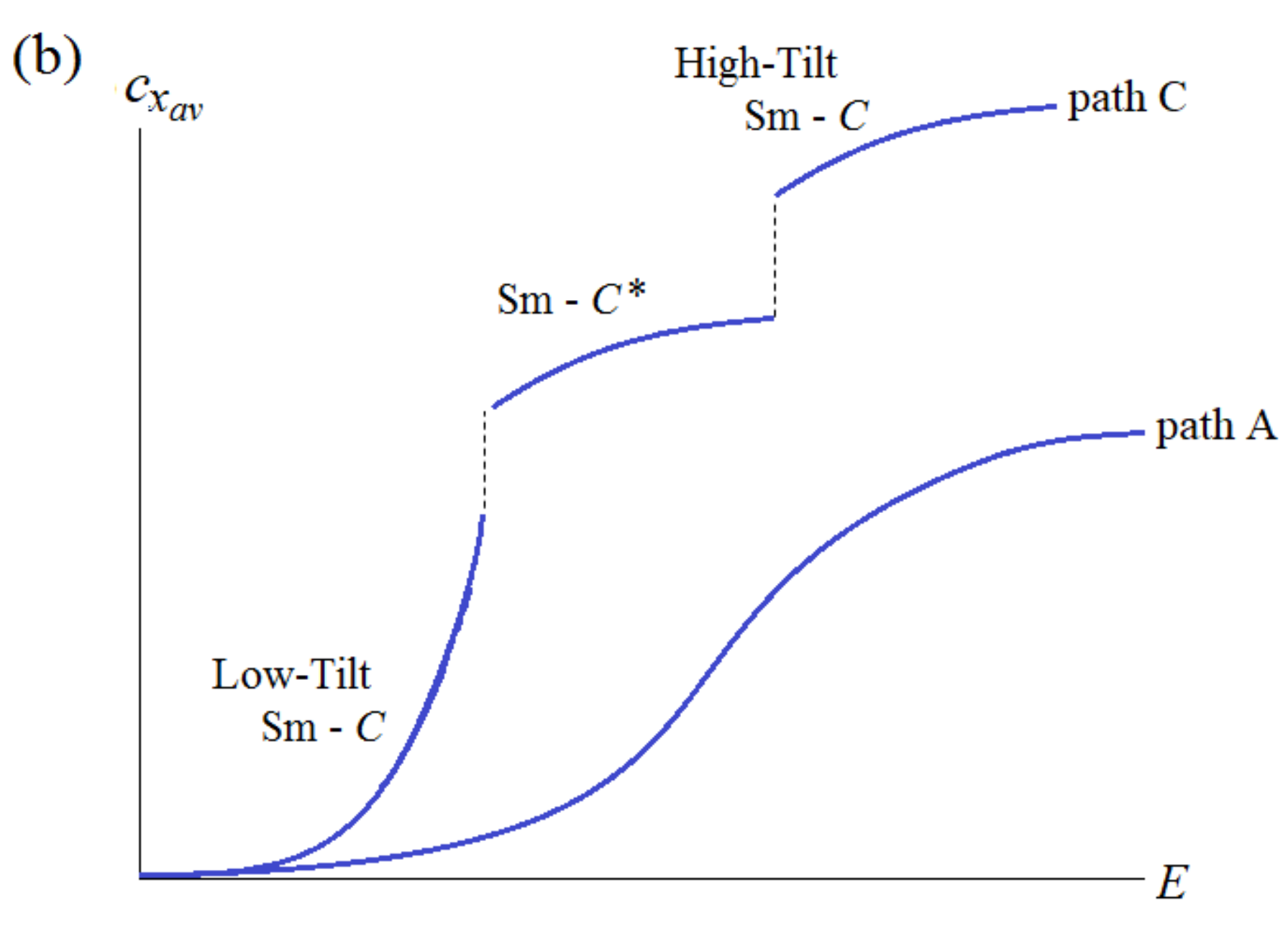}
\caption{Summary of electroclininic effect, i.e., average tilt ($c_{x_{\mathrm{av}}}$) vs electric field ($E$) for (a) Type I and (b) Type II. The corresponding fixed $T$ paths are shown in Fig.~\ref{Summary of E-T phase diagrams}(a) and (b). 
\label{Summary of tilt along paths}}
\end{figure}

Figure~\ref{Summary of tilt along paths} shows the behavior of the electroclinic effect when the field is ramped up from the Sm-$A$ phase. Fig.~\ref{Summary of tilt along paths}(a) shows that for Type I systems the tilt increases continuously with field, for $T>T_c$ (path A). For $T<T_c$ (path B) the tilt jumps at the low-tilt--high-tilt Sm-$C$--Sm-$C$ transition, and the system is bistable. For Type II systems (Fig.~\ref{Summary of tilt along paths}(b)) at high temperatures (path A) the tilt also increases continuously with field, while at lower temperatures (path C) the tilt exhibits {\em two} discontinuities, one at the Sm-$C$--Sm-$C^*$ transition, and the other at the Sm-$C^*$--Sm-$C$ transition, and the system is {\em tristable}. Of course, the tilt in the Sm-$C^*$ phase is modulated, but has a non-zero average (over one modulation period). The pair of tricritical points mean that along path H the Sm-$C$--Sm-$C^*$--Sm-$C$ transition sequence is continuous. For a path between the two tricritical points (Path I) the Sm-$C$--Sm-$C^*$--Sm-$C$ transition sequence has one continuous transition and one discontinuous transition. 

Interestingly, at first sight the the phase diagram for the compound CE8, proposed by Ghoddoussi et. al. \cite{Ghoddoussi}, resembles the Type I phase diagram of Fig.~\ref{Summary of E-T phase diagrams}(a). However, there are some important, fundamental differences. For example, the transition on the uniform Sm-$C$--Sm-$C$ phase boundary for CE8 is continuous (and without a critical point) whereas that of Fig.~\ref{Summary of E-T phase diagrams}(a) is discontinuous and does have a critical point. Additionally, the sign of the curvature of the phase boundary at zero-field differs for each system. The fundamental reason for these differences is that the compound CE8 exhibits a {\it continuous} zero field Sm-$A$--Sm-$C^*$ transition, whereas the phase diagram of Fig.~\ref{Summary of E-T phase diagrams}(a) only applies to systems with a discontinuous zero field Sm-$A$--Sm-$C^*$ transition.

There is however another theoretical model \cite{Belitz} that does predict phase diagrams with striking similarities to Figs.~\ref{Summary of E-T phase diagrams}(a) and (b), namely a model for a quantum system capable of displaying ferromagnetic (FM), antiferromagnetic (AFM) and paramagnetic (PM) phases. Loosely speaking, the AFM phase is analogous to the modulated, Sm-$C^*$ phase, while the FM and PM phases are analogous to the low and high tilt uniform Sm-$C$ phases. In some regions of parameter space, the quantum phase diagram exhibits a FM-PM quantum critical point, and a FM-PM-AFM quantum triple point, like our Type I system. In other regions of parameter space the quantum critical and triple points are replaced by a region of AFM phase, with a {\em single} tricritical point. In future work, we will investigate the analogies between the mesoscopic smectic and quantum systems, but in this article we focus on the smectic system.

The remainder of the paper is organized as follows. In Section \ref{sec:ModelFreeEnergy} we establish the model and free energy. In Section \ref{sec:EstimatingLengthScaleRatio} we estimate the length scale ratio that delineates Type I and Type II behavior, and also discuss ways to experimentally access each type of behavior. In Section \ref{sec:CommonFeatures}, we map out the small $E$ and low $T$ phase boundaries. For both Type I and Type II systems these parts of phase boundaries are similar. We analyze the most interesting region of the phase diagrams, near the uniform low tilt -- high tilt critical point in Sections \ref{sec:DeterminationUniformModulatedPhaseBndyTypeII} and \ref{sec:MappingModulatedPhaseBndyTypeIMUTriplePoint}, mapping out the Type II region in Section \ref{sec:DeterminationUniformModulatedPhaseBndyTypeII} and the Type I region in Section  \ref{sec:MappingModulatedPhaseBndyTypeIMUTriplePoint}. We discuss the crossover between Type I and Type II phase diagrams in Section \ref{sec:CrossoverBetweenTypeITypeII}. Numerical results are presented in Section \ref{sec:NumericalAnalysis}. In Section \ref{sec:Mapping ExperimentTETheoryRepsilon} we outline how experimental results could be compared with our theoretical predictions, and we conclude with some discussion of future plans in Section \ref{sec:Conclusion}.

\section{Model and Free Energy}
\label{sec:ModelFreeEnergy}

As shown in Fig.~\ref{Smectic Schematic} the order parameter for the tilted phases (either the uniform Sm-$C$ phase or the modulated Sm-$C^*$ phase) is the two-component vector, $\bf c (r)$ that describes the local tilt of the molecular director relative to the layers. The vector $\bf c $ is perpendicular to the layer normal $\bf\hat{z}$. The molecular director $\bf \hat n$ can be expressed in terms of $\bf c $ as follows:
\begin{eqnarray}
{\bf \hat n}= {\bf c} + \sqrt{1-|{\bf c}|^2} {\bf \hat z} \approx {\bf c} + {\bf \hat z}\;,
\label{n ito c}
\end{eqnarray}
where the simplifying approximation is valid for small tilt. In the untilted Sm-$A$ phase ${\bf c}=0$. In the uniform Sm-$C$ phase, ${\bf c (r)}=\bf c$, independent of position $\bf r$. In the modulated Sm-$C^*$ phase, the precession of the optical axis around the layer normal ${\bf \hat z}$ (shown in Fig.~\ref{Helical Sm-C* Phase}) corresponds to periodic, $z$-dependent ${\bf c (r)}={\bf c}(z)$. 

Our analysis is restricted to the ground states of the different phases, and does not consider thermal fluctuations of the order parameter ${\bf c (r)}$ or the layers. Previous analyses of the {\it non-chiral} Sm-$A$--Sm-$C$ transition have included fluctuation effects due to both layers and molecular tilt\cite{G&P,SaundersTCP}. Interestingly, at a Sm-$A$--Sm-$C$ tricritical point (where a first order phase boundary meets a second order phase boundary) the stronger, tricritical fluctuations of the molecular tilt can be shown to destroy the smectic layering\cite{SaundersTCP}. An analysis of the fluctuation effects near the {\it chiral} Sm-$A$--Sm-$C^*$ transition will be carried out at a later date \cite{Saunders unpublished}.

Having established the order parameter $\bf c (r)$, we employ standard Landau theory, expanding the free energy in powers of the order parameter and its spatial gradients. It is useful to start with the free energy for a zero-field, non-chiral system. One can subsequently incorporate chirality and non-zero field into the model through the inclusion of extra terms in the free energy. The Landau free energy for a zero-field, non-chiral system is
\begin{eqnarray}
f_\text{non-chiral}=f_\text{MF}({\bf c}) + f_\text{grad}({\nabla \bf c}) \;,
\label{f non chiral}
\end{eqnarray}
where the mean field part is given by
\begin{eqnarray}
f_\text{MF}=\frac{r(T)}{2}|{\bf c}|^2+\frac{u}{4}|{\bf c}|^4+\frac{v}{6}|{\bf c}|^6 \;,
\label{f MF}
\end{eqnarray}
with $r(T)=\alpha(T-T_0)$ the temperature dependent parameter that drives the transition. The constants $\alpha$ and $T_0$ are system dependent. In Section \ref{sec:Mapping ExperimentTETheoryRepsilon} we will detail the procedure for experimentally mapping from $T$ to $r(T)$, but for now it is enough to know that $r(T)$ is a monotonically increasing function of temperature. Both $u$ and $v$ temperature independent constants, and stability of the system requires that $v>0$. 

The gradient part is given by
\begin{eqnarray}
f_\text{grad}=\frac{K_{s}}{2}\left({\bf \nabla_\perp \cdot c}\right)^2+\frac{K_{b}}{2}\left| \partial_z {\bf c}\right|^2+\frac{K_t}{2}| \left({\bf \nabla_\perp \times c}\right)^2.
\label{f grad}
\end{eqnarray}
The elastic constants $K_{s}$, $K_{b}$ and $K_{t}$ are the standard constants for splay, bend and twist distortions of the molecular director ${\bf \hat n}$. The above expression for 
$f_\text{grad}$ is obtained by inserting ${\bf \hat n} \approx {\bf c} + {\bf \hat z}$ into the standard Frank elastic energy density \cite{deGennes and Prost}.

The above non-chiral free energy density $f_\textrm{non-chiral}$ is invariant under chiral transformations, e.g., switching from a left-handed to right-handed coordinate system. To generalize the free energy density to model the chiral Sm-$A$--Sm-$C^*$ transition, we add two terms to  $f_\textrm{non-chiral}$,  
\begin{eqnarray}
f = f_\text{non-chiral} + e\eta \left({\bf c \cdot }\left(\bf{ \nabla \times c}\right)\right) +eg {\bf \hat z}\cdot({\bf E} \times {\bf c}) \;.
\label{f chiral and field 1}
\end{eqnarray}
The first added term breaks chiral symmetry, and favors a helical modulation of $\bf c$ with pitch $P_0=\frac{2\pi}{q_0}=\frac{2\pi K_b}{e\eta}$. The second added term is responsible for the well-known electroclinic effect\cite{Meyer, Garoff and Meyer}, whereby a coupling between the tilt and polarization of the molecules allows an electric field, e.g.,  ${\bf E} = E {\bf \hat y}$, to break the inversion symmetry within the plane of the layers, and thus causing the molecules to uniformly tilt along a direction perpendicular to the field, i.e., ${\bf c}=c{\bf \hat x}$ \cite{polarization footnote}. The constant $g$ is proportional to the electric susceptibility. Thus the two chiral terms compete, with the first favoring a modulated state and the second favoring a uniform state. We include a common, dimensionless factor, $e$, which represents the degree of chirality, and is thus a monotonically increasing function of enantiomeric excess. In a racemate, $e=0$ and neither of the chiral terms is present. 

Taking ${\bf E} = E {\bf \hat y}$ our starting free energy is
\begin{eqnarray}
F= A_\perp\int_{0}^{L_z} dz\bigg[\frac{r(T)}{2}|{\bf c}|^2+\frac{u}{4}|{\bf c}|^4+\frac{v}{6}|{\bf c}|^6 + \frac{K_{b}}{2}\left(\partial_z {\bf c}\right)^2  - e \eta (c_x\partial_z c_y-c_y\partial_z c_x)- e g E c_x\bigg] \;,
\label{F}
\end{eqnarray}
where $A_\perp=\int dx dy$, i.e., the area of the system's layers, and $L_z$ is the length of the system along $z$. Since ${\bf c}$ is periodic along $z$, the integral can be converted into one over a single period $2\pi/q$, 
\begin{eqnarray}
 \int_{0}^{L_z}dz = L_z\frac{q}{2\pi}\int_0^{\frac{2\pi}{q}} dz\;,
\label{integral}
\end{eqnarray}
where we assume a sufficiently large $L_z$ so that the system accommodates an integral number of modulation periods. Identifying the volume $V=A_\perp L_z$, the free energy per unit volume is then
\begin{eqnarray}
\frac{F}{V}=\frac{q}{2\pi}\int_0^{\frac{2\pi}{q}} dz\bigg[\frac{r(T)}{2}|{\bf c}|^2+\frac{u}{4}|{\bf c}|^4+\frac{v}{6}|{\bf c}|^6 + \frac{K_{b}}{2}\left(\partial_z {\bf c}\right)^2  - e \eta (c_x\partial_z c_y-c_y\partial_z c_x)- e g E c_x\bigg] \;.
\label{F/V}
\end{eqnarray}

\section{Estimating the Length Scale Ratio and Experimental Parameters that Determine Type I vs Type II Behavior}
\label{sec:EstimatingLengthScaleRatio}

As discussed in Section \ref{sec:SummaryResults}, we find that systems with a first order Sm-$A$--Sm-$C^*$ transition can be categorized as Type I or Type II. Type I systems have a first order uniform low-tilt uniform high-tilt (Sm-$C$--Sm-$C$) phase boundary that terminates in an Ising-type critical point. In Type II systems the critical point is absent and there is a modulated (Sm-$C^*$) phase between the uniform low-tilt and uniform high-tilt phases. 

Whether the system is Type I or Type II depends on a ratio of length scales. We calculate this ratio more carefully using an instability analysis in Section \ref{sec:ModelTypeIIBehaviorCriticalLengthScaleRatio}, but here we use a heuristic energetic argument that better illustrates {\em why} there are distinct Type I and Type II systems, and also results in an estimate that is accurate to a factor of order one.

To determine whether a system is Type I we employ a length scale estimate to determine whether the uniform, tilted phase (Sm-$C$) is energetically favorable compared to the modulated, tilted state. As discussed in the above section, for ${\bf E}=E{\bf \hat y}$, the lowest energy uniform state has ${\bf c}=c_u{\bf \hat x}$. The equation of state for $c_u$ is found by minimizing the free energy of Eq.~(\ref{F/V}), which gives 
\begin{eqnarray}
E=(r c_u + u c_u^3 +v c_u^5)/eg\;.
\label{E c_u}
\end{eqnarray}
The energetic {\em cost} of a deviation away from this uniform state can be estimated by inserting a non-uniform ${\bf c}=c_u{\bf \hat x} +\sigma_y(z){\bf \hat y}$ into the the free energy of Eq.~(\ref{F/V}), in which $\eta=0$. By setting $\eta=0$ we intentionally omit the energetic gain for a chiral deviation $\sigma_y(z)$. To $\mathcal{O}(\sigma_y^2)$, the energetic cost is 
\begin{eqnarray}
\Delta f_{\sigma}=\frac{1}{2}\chi^{-1}_y\sigma_y^2 + \frac{1}{2}K_b(\partial_z \sigma_y)^2\;, 
\label{sigma_y energy}
\end{eqnarray}
where the susceptibility $\chi_y$ is given by
\begin{eqnarray}
\chi_y^{-1}=r+u c_u^2+v c_u^4=\frac{egE}{c_u}\;.
\label{susceptibilty}
\end{eqnarray}
The second equality is obtained using the equation of state Eq.~(\ref{E c_u}). Equation~(\ref{sigma_y energy}) then gives a correlation length which can be interpreted as the length scale above which deviations away from the uniform ground state come at a significant energetic cost:
\begin{eqnarray}
\xi_y=\sqrt{\frac{K_b}{\chi_y^{-1}}}=\sqrt{\frac{K_b c_u}{egE}}=\sqrt{\frac{K_b}{(r+u c_u^2+v c_u^4)}}\;, 
\label{xi_y}
\end{eqnarray}
Now we remind ourselves that the $\eta\neq0$ term favors a chiral modulation along $z$ with a length scale $P_0=\frac{2\pi}{q_0}=\frac{2\pi K_b}{e\eta}$, and argue (heuristically) that such a modulation will {\em not} be energetically favorable if it occurs on a length scale large compared to $\xi_y$. In other words, the uniform state is favorable if $\frac{\xi_y}{P_0}<1$. To determine when a system is Type I, we apply this criterion at the uniform low-tilt uniform high-tilt (Sm-$C$--Sm-$C$) critical point. The location of this critical point in $r$-$E$ space is found by solving:
\begin{eqnarray}
\frac{dE}{dc_u}=0,\ \frac{d^2E}{dc_u^2}=0\;, 
\label{critical point condition}
\end{eqnarray}
with $E(c_u)$ given by Eq.~(\ref{E c_u}). Solving the three equations in Eqs.~(\ref{E c_u}) and (\ref{critical point condition}) yields the critical point values:
\begin{eqnarray}
c_{u_c}&=& \sqrt{\frac{-3u}{10v}},\nonumber\\
r_c&=&\frac{9u^2}{20v}, \nonumber\\
E_c&=&\frac{6u^2c_{u_c}}{25egv} \;,
\label{critical point values}
\end{eqnarray}
so that at the critical point the correlation length is
\begin{eqnarray}
\xi_{y_c}=\sqrt{\frac{25K_bv}{6u^2}}\;.
\label{xi_y_c}
\end{eqnarray}
Thus the length scale ratio that determines whether the system is Type I or Type II is
\begin{eqnarray}
\frac{\xi_{y_c}}{P_0}\sim e\eta\sqrt{\frac{v}{K_bu^2}}\;.
\label{m}
\end{eqnarray}
where we have omitted constants of order one.

We note that the sign of the length scale is determined by the sign of the chirality, and simply corresponds to the handedness of the chiral modulation. For $\frac{\xi_{y_c}}{P_0}<1$ the chiral modulations are too energetically costly at the critical point, and the system will be Type I, while for $\frac{\xi_{y_c}}{P_0}>1$ they are energetically favorable and the system will be Type II.

That $\frac{\xi_{y_c}}{P_0}$ is a monotonically increasing function of enantiomeric excess $e$ makes sense, since modulated states are favored in systems with larger enantiomeric excess. Moreover, it suggests an experimental means for tuning a system from Type I to Type II behavior, i.e., by doping a low chirality Sm-$C^*$ system with a high chirality, or tight-pitch Sm-$C^*$ compound\cite{footnote on antiferroelectric or orthoconic Sm Cs}. Another way to tune a system from Type I to Type II behavior would be to reduce the magnitude of $u$, i.e., to reduce the strength of the first order Sm-$A$--Sm-$C^*$ transition, or to use a compound that has a first order Sm-$A$--Sm-$C^*$ close to tricriticality. It has been established\cite{Roberts} that de Vries materials have transitions close to tricriticality, so we suggest that by doping a weakly first order de Vries system with a high chirality compound, one could tune from Type I to Type II behavior. 

\begin{figure}[ht]
\includegraphics[scale=0.6]{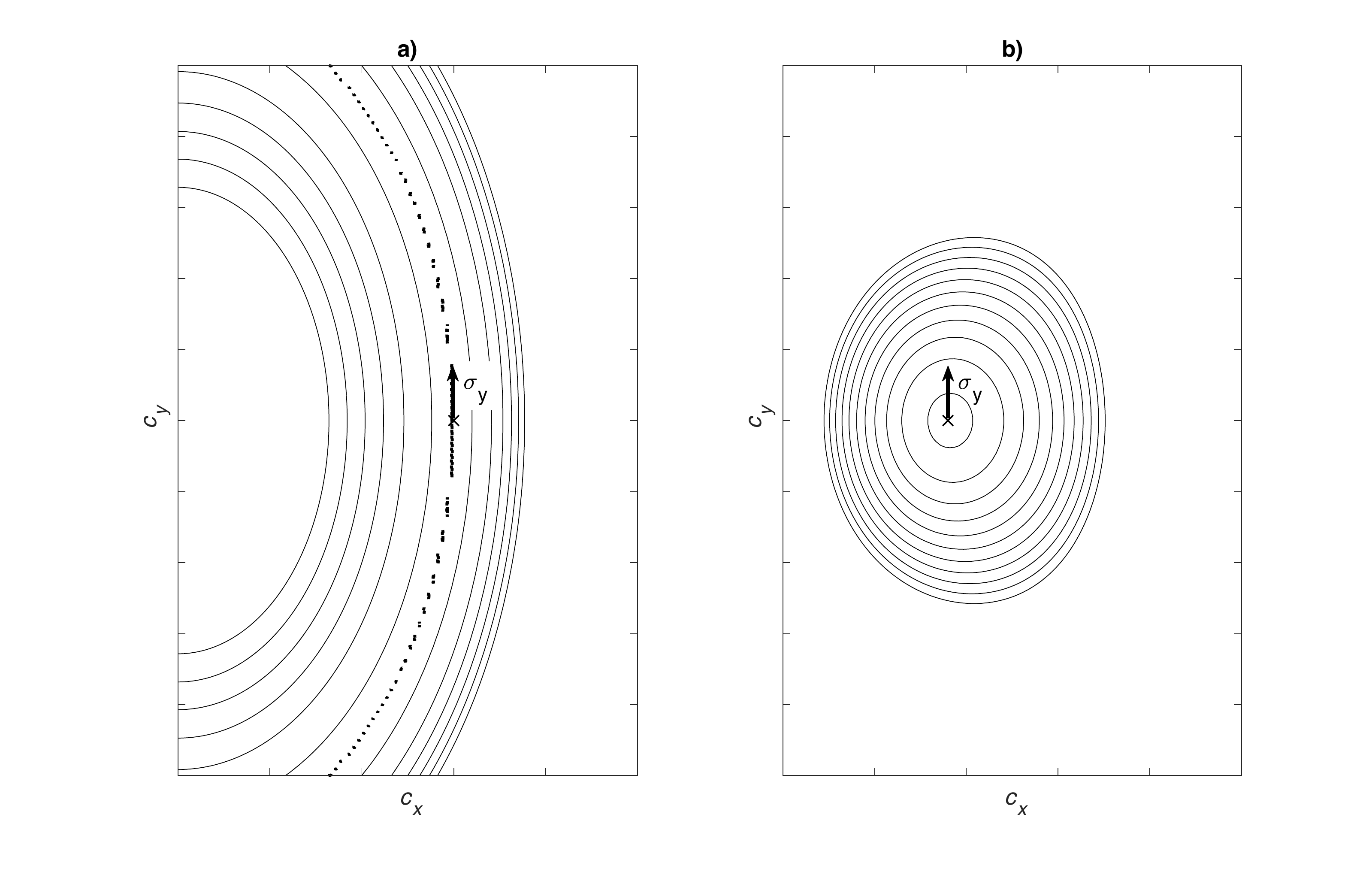}
\caption{(a) Contours of the energy landscape in the ordered phase near a zero field Heisenberg critical point, for systems with $u\geq0$. The dotted contour corresponds to the energetic minimum. A transverse deviation $\sigma_y$ from the state ``x" keeps the system in the energetic minimum and thus occurs at no energetic cost. (b) Contours of the energy landscape in the ordered phase near a finite field Ising critical point. The energy minimum is shown with an ``x."  A transverse deviation $\sigma_y$ from the state ``x" requires a finite energy cost.
\label{EnergyLandscape}}
\end{figure}

From the expression for $m$ in Eq.~(\ref{m}) we also see that for a system with a tricritical Sm-$A$--Sm-$C^*$, i.e., with $u=0$, the ratio $\frac{\xi_{y_c}}{P_0}$ will diverge, and Type II behavior will always result. However, such a system will {\em not} have a first order uniform low-tilt uniform high-tilt (Sm-$C$--Sm-$C$) phase boundary, nor the associated Ising type critical point. If we again consider a model with $\eta=0$, i.e., one in which we intentionally omit the energetic gain for a chiral deviation, any system with $u\geq0$ will have a Heisenberg type critical point at $r=E=0$. Since $E=0$, no special direction is picked out for $\bf c$, i.e., the symmetry is {\em spontaneously} broken at the continuous transition to the Sm-$C$ phase. The associated Goldstone mode means that deviations away from the spontaneously chosen direction of $\bf c$ cost no energy, so that an infinitesimal amount of chirality will result in a modulated state, as shown in Fig.~\ref{EnergyLandscape}(a). This differs from the  $u<0$ critical point at $E_c\neq 0$, which is Ising like and, as shown in Fig.~\ref{EnergyLandscape}(b), does not have a Goldstone mode, meaning that the chirality must exceed a finite threshold for a modulated state to be energetically favorable.

\begin{figure}[ht]
\includegraphics[scale=0.5]{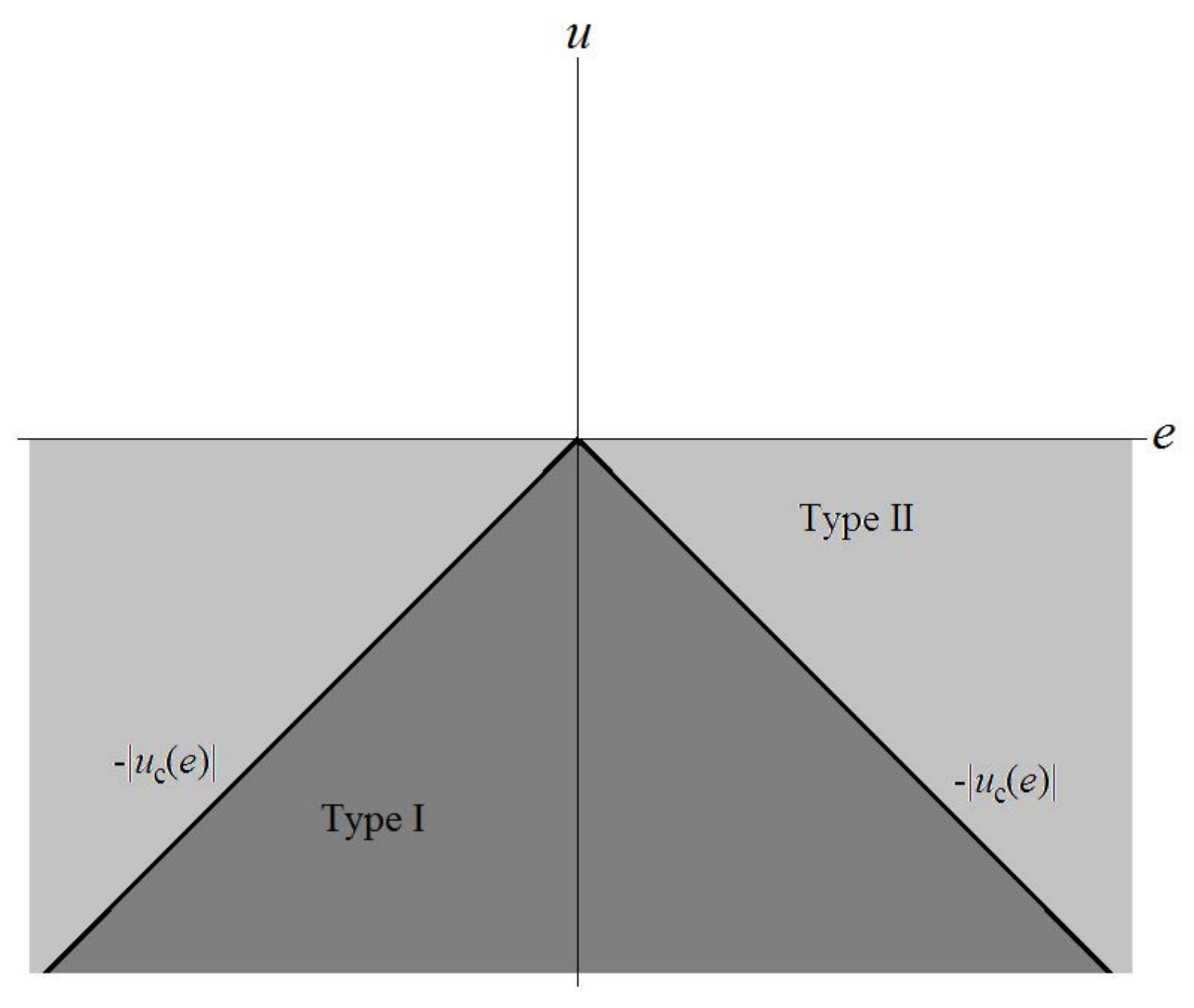}
\caption{A classification diagram for the different classes of chiral Sm-$C$ systems. The class of system depends on whether the Sm-$A$--Sm-$C^*$ transition is continuous ($u>0$) or first order ($u<0$). If $u<0$, the system can be Type II (small $\lvert u \rvert < \lvert u_c \rvert$) or Type I (large $\lvert u \rvert > \lvert u_c \rvert$).
\label{u-eClassificationDiagram}}
\end{figure}

\begin{figure}[ht]
\includegraphics[scale=0.5]{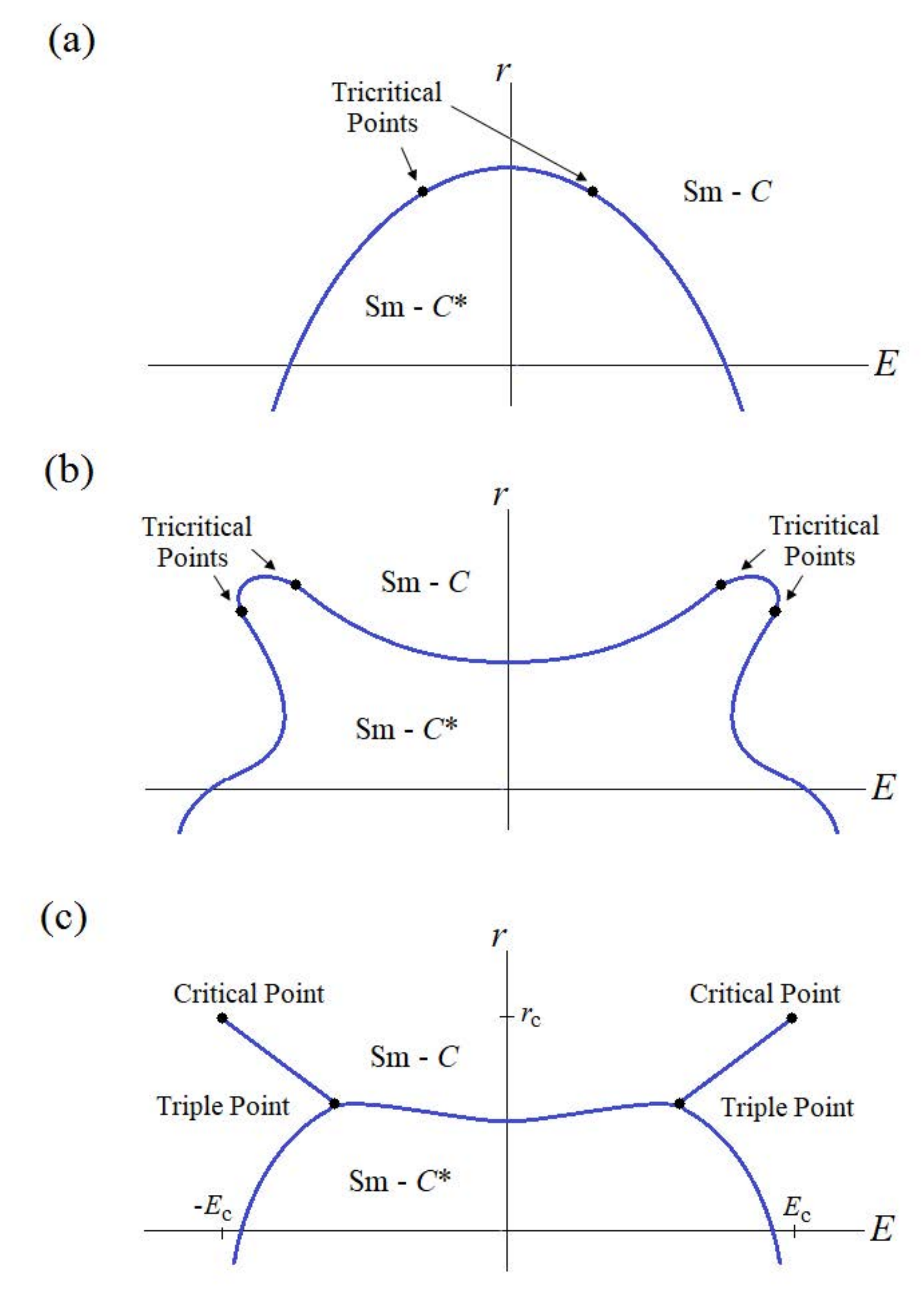}
\caption{Evolution of the phase diagram with decreasing $u$. (a) System with $u>0$, i.e., continuous Sm-$A$--Sm-$C^*$.  (b) Type II system with $-\lvert u_c \rvert<u<0$, e.g., a weakly first order Sm-$A$--Sm-$C^*$. (c) Type I system with $u<-\lvert u_c \rvert$, e.g., a strongly first order Sm-$A$--Sm-$C^*$.
\label{DecreasingUPhaseDiagrams}}
\end{figure}

Figure~\ref{u-eClassificationDiagram} shows a classification diagram for the different classes of chiral Sm-$C$ systems. The class of system depends on whether the Sm-$A$--Sm-$C^*$ transition is continuous ($u>0$) or first order ($u<0$). If $u<0$, the system can be Type II (small $\lvert u \rvert < \lvert u_c \rvert$) or Type I (large $\lvert u \rvert > \lvert u_c \rvert$), where
\begin{eqnarray}
\lvert u_c \rvert\sim \lvert e \rvert \eta \sqrt{\frac{v}{K_B}}\;,
\label{uc}
\end{eqnarray}
and again we omit constants of order one. If one were to ramp $u$ up from $u < u_c<0$ to $u>0$ one would see an interesting evolution of phase diagrams shown in Fig.~\ref{DecreasingUPhaseDiagrams}. For $u<u_c$, the phase diagram  has a triple point and a critical point as shown in Fig.~\ref{DecreasingUPhaseDiagrams}(c). As $u$ is raised, the triple point approaches the critical point, and at $u=u_c$ the two points coalesce into what we term a triple critical point. Then for $u_c<u<0$ the triple critical point splits into two tricritical points connected by a second order phase boundary, as shown in Fig.~\ref{DecreasingUPhaseDiagrams}(b). Experimentally, variation of $u$ could perhaps be achieved by varying concentrations of a mixture of $u>0$ and $u<0$ compounds\cite{TricriticalExperimentRatna}. 

%

Having discussed the delineation of Type I vs Type II behavior in terms of experimental parameters, it is useful to now rescale our model. Doing so allows us to work with a non-dimensionalized model with only three parameters. We rescale by letting:
\begin{eqnarray}
{\bf c}&=&c_{u_c}{\bf C},\nonumber\\
r(T)&=&r_cR(T), \nonumber\\
E&=&E_c \epsilon, \nonumber\\
z&=&\frac{Z}{q_0}\;.
\label{resclaing}
\end{eqnarray}
${\bf C}$ is the rescaled tilt, $R(T)$ is the rescaled temperature dependent parameter, $\epsilon$ is a dimensionless field, and $Z$ is the dimensionless length scale. Equation~(\ref{critical point values}) gives $c_{u_c}$, $r_c$ and $E_c$, and the zero field helical wavevector $q_0=e\eta/K_B$. Inserting the above rescalings into Eq.~(\ref{F/V}) gives the following rescaled free energy density:
\begin{eqnarray}
\frac{F}{V}=\frac{q}{2\pi q_0}\bigg(\frac{-27u^3}{200 v^2}\bigg)\int_0^{\frac{2\pi q_0}{q}} dZ\bigg[\frac{R(T)}{2}|{\bf C}|^2&-&\frac{1}{6}|{\bf C}|^4+\frac{1}{30}|{\bf C}|^6 + \frac{m^2}{15}\bigg( \left(\partial_Z {\bf C}\right)^2  - 2(C_x\partial_Z C_y-C_y\partial_Z C_x) \bigg) \nonumber\\
&-& \frac{8}{15} \epsilon C_x\bigg] \;,
\label{F/V rescaled}
\end{eqnarray}
where 
\begin{eqnarray}
m\equiv\pm\frac{e\eta}{u}\sqrt{\frac{50v}{3K_b}}\sim\frac{\xi_{y_c}}{P_0}\;,
\label{ec}
\end{eqnarray}
is equivalent (up to a constant of order one) to the length scale ratio $\frac{\xi_{y_c}}{P_0}$ that determines Type I versus Type II behavior. In Section \ref{sec:DeterminationUniformModulatedPhaseBndyTypeII} we will use an instability analysis to show that for $\lvert m \rvert <1$ the system is Type I while for $\lvert m \rvert >1$ it is Type II. We note that since $u<0$, the overall coefficient is positive. 

The above rescaled free energy density only has three parameters: $R(T)$, $\epsilon$, $m$. For a given system (characterized by $m$) we can then map out the phase diagram in $R(T)$-$\epsilon$ space. In Section \ref{sec:Mapping ExperimentTETheoryRepsilon} we outline the way in which one can convert an experimental phase diagram from $T$-$E$ space to $R(T)$-$\epsilon$ space, and also how one can determine the experimental $m$ number that characterizes the system. 

\section{Common Features of Type I and Type II Phase Diagrams}
\label{sec:CommonFeatures}

From the Type I and Type II phase diagrams shown in Fig.~\ref{Summary of E-T phase diagrams} we see that the uniform-modulated phase boundaries are qualitatively similar in two regions: {\bf(A)} for temperatures far below the Sm-$A$--Sm-$C^*$ transition temperature, i.e., $T\ll T_{AC^*}$  and {\bf(B)} for small electric fields, i.e., $E\approx0$. We begin our analysis by mapping out the phase boundary in these two regions. 

\subsection{Uniform-Modulated Phase Boundary Far Below the Sm-$A$--Sm-$C^*$ Transition Temperature}
\label{sec:UniformModulatedPhaseBoundaryFarBelow}

Deep within the Sm-$C^*$ phase, i.e., far below the Sm-$A$--Sm-$C^*$ transition temperature, ramping up the electric field ($\epsilon$) will cause the helical modulation to unwind. Eventually, at finite critical field $\epsilon_{q\rightarrow0}$ the pitch diverges, or equivalently, the modulation wave vector $q$ vanishes. Determining $\epsilon_{q\rightarrow0}(R)$ thus gives the location of the Sm-$C$--Sm-$C^*$ (uniform-modulated) phase boundary. The calculation of $\epsilon_{q\rightarrow0}(R)$ for systems with $u<0$ is not very different from those with $u>0$. The reason is that deep within the Sm-$C^*$ phase there is only a single energetic minimum for the uniform phase regardless of the sign of $u$. This is {\em not} the case for temperatures near or above the Sm-$A$--$Sm-C^*$ transition temperature, which leads to significantly different and more interesting behavior for systems with $u<0$, e.g., the possibility of Type I or Type II systems, and reentrance.

To calculate $\epsilon_{q\rightarrow0}(R)$ we follow a similar procedure to that outlined in Refs.~\onlinecite{Schaub and Mukamel} and \onlinecite{deGennes and Prost}, and start by assuming a helical modulation with a uniform tilt magnitude, i.e., 
\begin{eqnarray}
C_x(Z)=C\cos(\phi(Z)) \;,\nonumber\\
C_y(Z)=C\sin(\phi(Z))\;,
\label{ansatz 1}
\end{eqnarray}
with $\partial_Z C =0$. Inserting this form of $\bf C$ into the free energy density (Eq.~(\ref{F/V rescaled})) yields
\begin{eqnarray}
\frac{F_{R\ll1}}{V}=\frac{q}{2\pi q_0}\bigg(\frac{-27u^3}{200 v^2}\bigg)\int_0^{\frac{2\pi q_0}{q}} dZ &\bigg[& \frac{R(T)}{2}C^2-\frac{1}{6}C^4+\frac{1}{30}C^6 + \nonumber \\ & & \frac{m^2C^2}{15}\bigg( \left(\partial_Z \phi \right)^2  - 2\partial_Z \phi  \bigg)- \frac{8}{15} \epsilon C \cos(\phi) \bigg],
\label{polar f}
\end{eqnarray}
where $R(T)\ll1$ corresponds to the system being deep in the Sm-$C^*$ phase. In the absence of an electric field, the above energy is minimized by $\phi=Z=q_0z$, which corresponds to a perfect helix with wavevector $q_0=\frac{e\eta}{K_b}$. For non-zero field, minimization of $F_{R\ll1}$ with respect to $\phi(Z)$ gives the following Sine-Gordon equation:
\begin{eqnarray}
\frac{d^2 \phi}{dZ^2}= \frac{4\epsilon}{m^2C}\sin\phi\;.
\label{EL phi}
\end{eqnarray}
Solving this equation, along with the condition $q\rightarrow 0$ gives an implicit equation for $\epsilon_{q\rightarrow0}$, the location of the Sm-$C$--Sm-$C^*$ (uniform-modulated) transition:
\begin{eqnarray}
C(\epsilon_{q\rightarrow0},R(T))=\frac{64}{\pi^2m^2}\epsilon_{q\rightarrow0}\;,
\label{Eq0}
\end{eqnarray}
with $C(\epsilon_{q\rightarrow0},R(T))$ the magnitude of the tilt in the uniform (Sm-$C$) phase which is found setting $\partial_Z \phi=0$ and $\phi=0$ in the free energy of Eq.~(\ref{polar f}), and minimizing the resulting free energy with respect to $C$. Doing so gives:
\begin{eqnarray}
R(T)C - \frac{2}{3}C^3+\frac{1}{5}C^5=\frac{8}{15}\epsilon\;.
\label{cU}
\end{eqnarray}
Combining Eqs.~(\ref{Eq0}) and (\ref{cU}) allows us to obtain an expression for $\epsilon_{q\rightarrow0}(R)$ or, equivalently $R_{q\rightarrow0}(\epsilon)$
\begin{eqnarray}
R_{q\rightarrow0}(\epsilon)=\frac{(\pi m)^2}{120}\bigg[1+\frac{5}{4}\bigg(\frac{8}{\pi m} \bigg)^6\epsilon^2-\frac{3}{8}\bigg(\frac{8}{\pi m} \bigg)^{10}\epsilon^4\bigg]
\;.
\label{rq0}
\end{eqnarray}
We note that despite the positive $\epsilon^2$ coefficient the curvature of the $R_{q\rightarrow0}(\epsilon)$ boundary is negative due to the $\epsilon^4$ term which dominates for large $\epsilon$, i.e., in the  $R_{q\rightarrow0}(\epsilon)\ll1$ region in which this unwinding treatment is valid.

\subsection{Uniform-Modulated Phase Boundary for Small Electric Field}
\label{sec:UniformModulatedPhaseBoundarySmallE}

\begin{figure}[ht]
\includegraphics[scale=0.5]{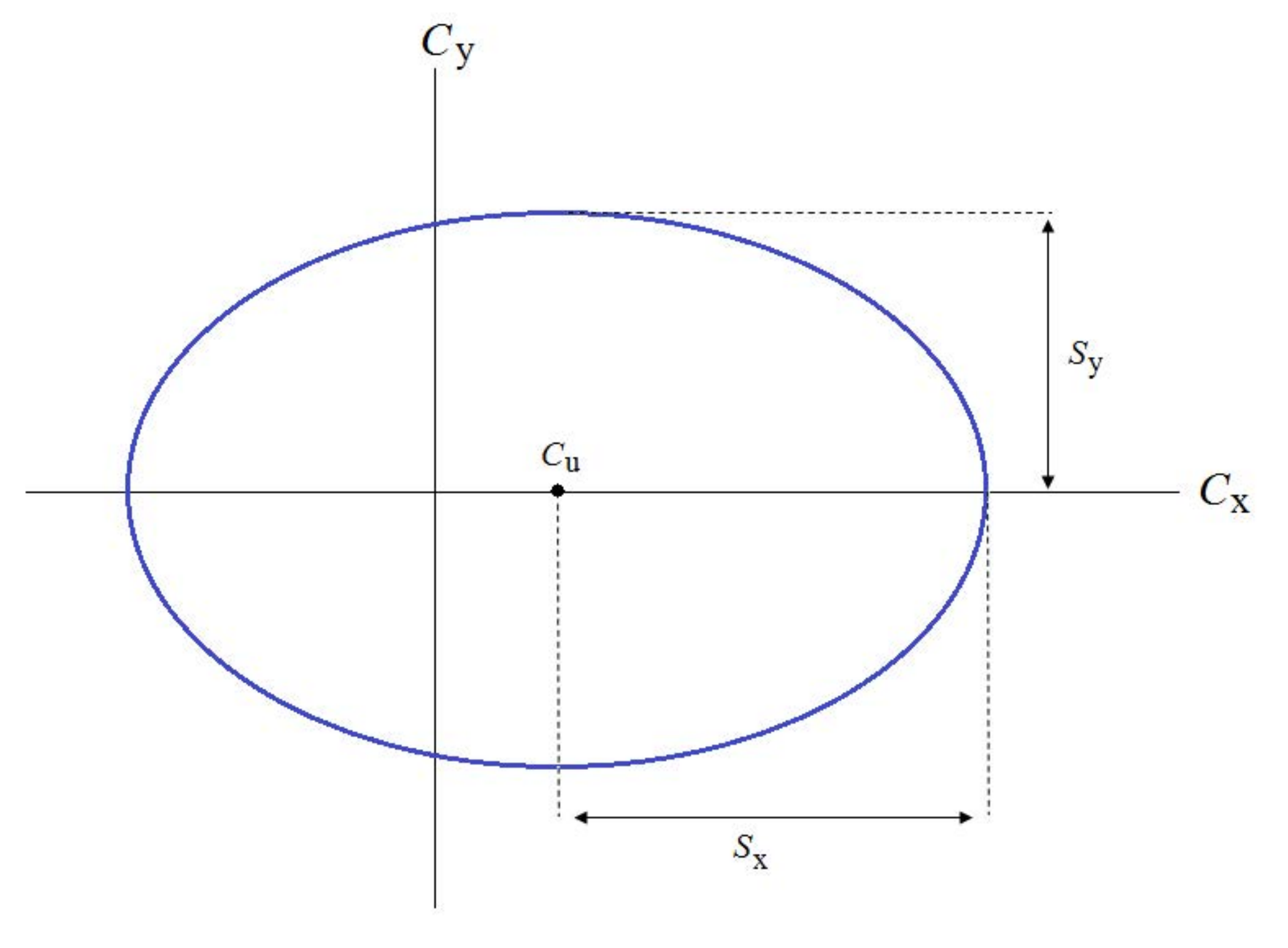}
\caption{Elliptical shape of helical modulation projected onto the $C_y-C_x$ plane. $C_u$ is the tilt averaged over one modulation period. $s_x$ and $s_y$ are the semi-major and semi-minor axes. 
\label{Small field Shape of Helical Modulation}}
\end{figure}

In the absence of an electric field the system will transition from the uniform Sm-$A$ phase to modulated Sm-$C^*$ as temperature is lowered. In the Sm-$C^*$ phase, the tilt director ${\bf c}(z)$ has a perfectly helical modulation with wave vector $q_0=\frac{e\eta}{K_b}$. For small finite field $\epsilon\approx 0$ the perfect helix will be distorted slightly as shown in Fig.~\ref{Small field Shape of Helical Modulation} and we approximate the tilt director ${\bf c}(z)$ with the following ansatz:
\begin{eqnarray}
C_x(Z)=C_u+s_x\sin\bigg(\frac{qZ}{q_0}\bigg) \;,\nonumber\\
C_y(Z)=s_y\cos\bigg(\frac{qZ}{q_0}\bigg)\;,
\label{ansatz 2}
\end{eqnarray}
where $C_u$ the magnitude of the tilt along $\bf \hat x$ in the high temperature uniform state, and $q$ is the wave-vector of the lowest energy chiral modulation. The positive amplitudes $s_x$ and $s_y$ describe the shape of the helical modulation, as shown in Fig.~\ref{Small field Shape of Helical Modulation}. For small electric field, $q\approx q_0$, $s_x\approx s_y$, and $c_u \approx 0$.

It is useful to express the modulated part of the ansatz in terms of a modulation order parameter $s$ that is nonzero in the modulated phase, regardless of the $x$ and $y$ weightings.
\begin{eqnarray}
C_x(Z)&=&C_u+s\cos(\alpha)\sin\bigg(\frac{qZ}{q_0}\bigg) =C_u+s\sigma\sin\bigg(\frac{qZ}{q_0}\bigg) \;,\nonumber\\
C_y(Z)&=&s\sin(\alpha)\cos\bigg(\frac{qZ}{q_0}\bigg)=s\sqrt{1-\sigma^2}\cos\bigg(\frac{qZ}{q_0}\bigg)\;,
\label{ansatz 2}
\end{eqnarray}
where $\sigma = \cos(\alpha)$ with $0\leq\alpha\leq\pi/2$, and thus $0\leq\sigma\leq1$. 

The process now is to insert the ansatz into the free energy, Eq.~(\ref{F/V rescaled}). Inserting the ansatz into the free energy, and performing the integral results in a free energy that depends on $C_u$, $s$, $\sigma$, and $q$. Minimizing with respect to $q$ yields
\begin{eqnarray}
q_{\min}=2q_0 \sigma \sqrt{1-\sigma^2}\;.
\label{q min}
\end{eqnarray}
Inserting $q=q_{\min}$ back into $F$, and minimizing with respect to $\sigma$ yields
\begin{eqnarray}
\sigma_{\min}=\frac{1}{\sqrt{2}}\sqrt{\frac{3s^4-10s^2+16m^2-24C_u^4+40C_u^2}{3s^4-10s^2+16m^2+18C_u^2s^2}}\;.
\label{sigma min}
\end{eqnarray}

At zero field, the uniform piece of $\bf C$ is zero, i.e., $C_u=0$, and Eqs.~(\ref{q min}) and (\ref{sigma min}) rightly yield $\sigma_{\min}=1/\sqrt{2}$ and $q_{\min}=q_0$. The remaining zero field free energy density is
\begin{eqnarray}
\frac{F_{\epsilon=0}}{V}=-\frac{27u^3}{800v^2}\bigg[\left(R-\frac{2}{15}m^2\right)s^2 -\frac{1}{6}s^4+\frac{1}{60}s^6\bigg]\;.
\label{F_E=0}
\end{eqnarray}

The negative $s^4$ coefficient means that the Sm-$A$--Sm-$C^*$ transition is first order, and upon entry to the Sm-$C^*$ phase the tilt magnitude jumps to a nonzero value $s_{1^\text{st}}^0$, where the superscript $0$ indicates the zero field result. The transition temperature $R_{1^\text{st}}^0\equiv R(T_{1^\text{st}}^0)$ can be found by equating the Sm-$A$ and Sm-$C^*$ minima of $F_{\epsilon=0}(s)$, i.e., solving the simultaneous equations
\begin{eqnarray}
\frac{dF_{\epsilon=0}}{ds_{1^\text{st}}^0}&=0& , \nonumber\\
F_{\epsilon=0}(s=0)&=&F_{\epsilon=0}(s_{1^\text{st}}^0)\; ,
\label{First order F_E=0}
\end{eqnarray}
for $R=R_{1^\text{st}}^0$ and $s_{1^\text{st}}^0$. This yields
\begin{eqnarray}
R_{1^\text{st}}^0 & =& \frac{5}{12}+\frac{2m^2}{15},\nonumber\\
s_{1^\text{st}}^0 & =& \sqrt{5}\;.
\label{r, s E=0}
\end{eqnarray}

\begin{figure}[ht]
\includegraphics[scale=0.5]{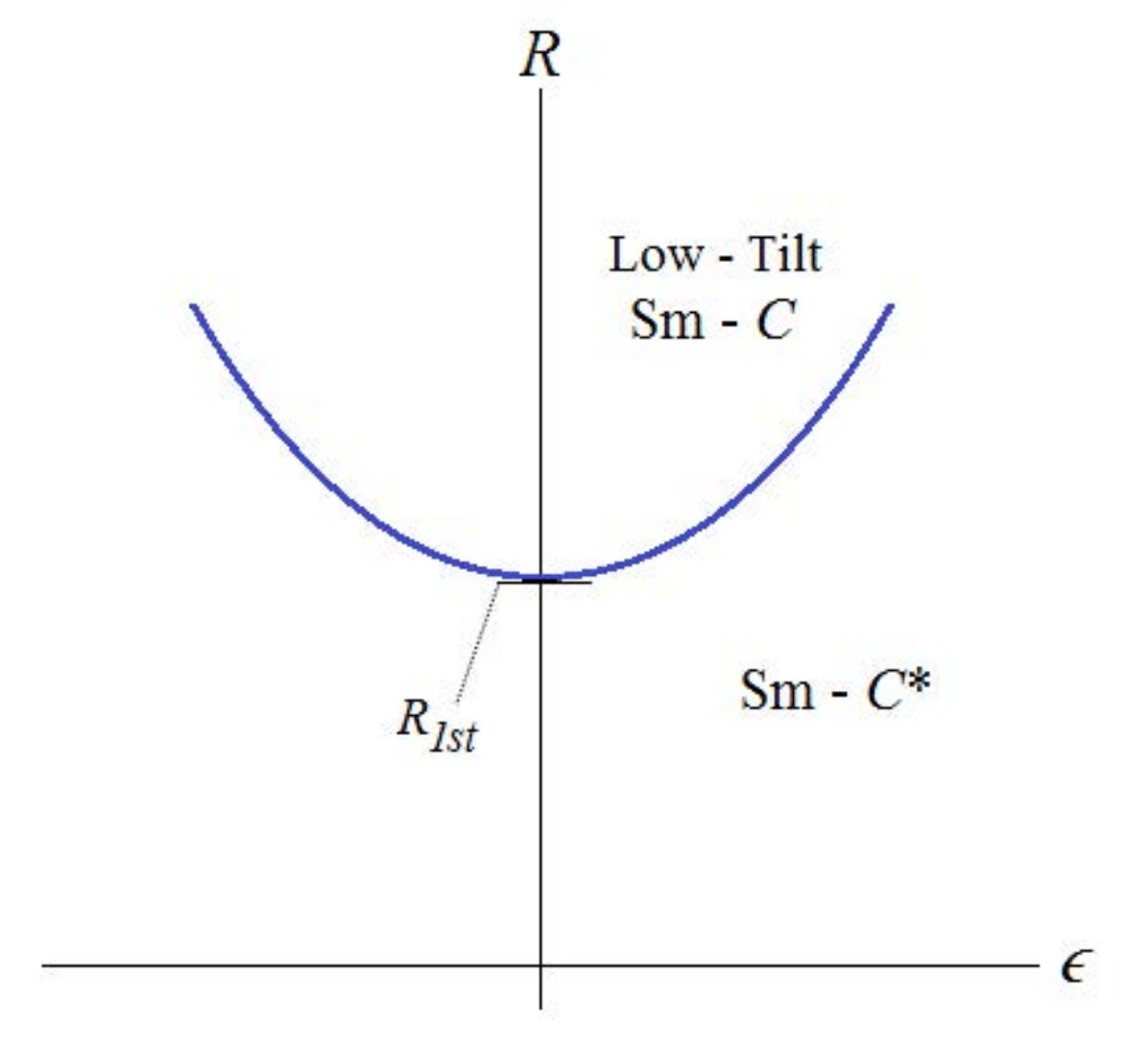}
\caption{The Sm-$C$--Sm-$C^*$ phase boundary in $R-\epsilon$ space near $\epsilon=0$ for  systems with a first order Sm-$A$--Sm-$C^*$ transition. Note that the curvature is positive which, as discussed in the text, means that the system displays a reentrant Sm-$C$--Sm-$C^*$--Sm-$C$ phase sequence as $\epsilon$ is increased. 
\label{uni-mod curvature}}
\end{figure}

For nonzero field $\epsilon\neq0$, $C_u\neq0$ and the system is no longer in the untilted, Sm-$A$ phase. Instead it is either in the uniform Sm-$C$ phase or the modulated Sm-$C^*$ phase. To determine the Sm-$C$--Sm-$C^*$ boundary near $\epsilon=0$, we expand the free energy in powers of $C_u$, which will be small for small $\epsilon$. Inserting $q=q_{\min}$ and $\sigma=\sigma_{\min}$ (given by Eqs.~(\ref{q min}) and (\ref{sigma min})) into the free energy and expanding to quadratic order in $C_u$ we find
\begin{eqnarray}
\frac{F_{\epsilon\approx0}(s,C_u)}{V}=\frac{F_{\epsilon=0}(s)}{V}-\frac{27u^3}{200v^2}\bigg[\beta(s)C_u^2-\frac{8}{15} \epsilon C_u\bigg]\;,
\label{F_E neq 0}
\end{eqnarray}
where 
\begin{eqnarray}
\beta(s)=\frac{R}{2}-\frac{1}{3}s^2+\frac{3}{40}s^4\;.
\label{beta}
\end{eqnarray}
The above free energy density has a minimum $C_u$:
\begin{eqnarray}
C_{u_{\min}}=\frac{4\epsilon}{15\beta(s)}\;.
\label{cu min}
\end{eqnarray}
Inserting $C_{u_{\min}}$ back into Eq.~(\ref{F_E neq 0}) results in a purely $s$ dependent free energy density
\begin{eqnarray}
\frac{F_{\epsilon\approx0}(s)}{V}=\frac{F_{\epsilon=0}(s)}{V}+\frac{6u^3}{625v^2}\frac{\epsilon^2}{\beta(s)}\;.
\label{F_E neq 0 1}
\end{eqnarray}
This time we equate the Sm-$C$ and Sm-$C^*$ minima of $F_{\epsilon\approx0}(s)$, i.e., solving the simultaneous equations
\begin{eqnarray}
\frac{dF_{\epsilon\approx0}}{ds_{1^\text{st}}(\epsilon)}&=0& \nonumber\\
F_{\epsilon\approx0}(s=0)&=&F_{\epsilon=0}(s_{1^\text{st}}(\epsilon)).\;
\label{First order F_E=0 1}
\end{eqnarray}
This results in a modified $s_{1^\text{st}}(\epsilon)=s_{1^\text{st}}^0+\mathcal{O}(\epsilon^2)$ and a Sm-$C$--Sm-$C^*$ boundary in $R-\epsilon$ space:
\begin{eqnarray}
R_{1^\text{st}}(\epsilon)&=& R_{1^\text{st}}^0+ \frac{8\epsilon^2} {15\beta(s_{1^{st}}^0)\beta(0))}+\mathcal{O}(\epsilon^4)\;.
\label{r for E not 0}
\end{eqnarray}
Noting that $\beta(s_{1^\text{st}}^0)>0$, and  $\beta(0)>0$, we see that the curvature of the 1st order Sm-$C$--Sm-$C^*$ phase boundary is positive, opposite to that of systems with a continuous Sm-$C$--Sm-$C^*$ phase transition. This means that, as shown in Fig.~\ref{uni-mod curvature}, if we start with a system in the Sm-$A$ phase (close to the Sm-$A$--Sm-$C^*$ transition temperature) and ramp up the field then the system will eventually jump from a uniform (Sm-$C$) to modulated (Sm-$C^*$) state. Of course, for sufficiently large fields the system must eventually transition back to a uniform (Sm-$C$) state. Thus, the system displays a reentrant Sm-$C$--Sm-$C^*$--Sm-$C$ phase sequence. 

\section{Determination of the Uniform--Modulated Phase Boundary for Type II Systems}
\label{sec:DeterminationUniformModulatedPhaseBndyTypeII}

Next we consider the most interesting region of the phase diagram, near the uniform low-tilt--high tilt critical point. We will show that there is a length scale ratio $m$ which determines whether the system is Type I or Type II. For $m>1$ the uniform phase at the critical point is unstable to modulation, and the system is Type II. In this case a Sm-$C$--Sm-$C^*$ phase boundary surrounds the {\em unstable} uniform critical point, as shown in Fig.~\ref{shrinking modulated region}. The modulated region shrinks as $m$ is reduced towards 1. For $m<1$ we show that the critical point is stable, and that there is a first order phase boundary between the uniform high and low tilt states. This first order line meets the modulated region at a triple point, where uniform low-tilt phase, the uniform high-tilt phase, and the modulated phase are all energetically equivalent. We map out the phase boundaries for Type I systems in Section \ref{sec:MappingModulatedPhaseBndyTypeIMUTriplePoint}.

\subsection{Model for Type II Behavior and Determination of the Critical Length Scale Ratio}
\label{sec:ModelTypeIIBehaviorCriticalLengthScaleRatio}

Since we are interested in the phase diagram near the critical point ($\epsilon_c=1$,$R(T_c)=1$), we expand ${\bf C}=C_u {\bf \hat x}$, $\epsilon$ and $R$ near their critical point values:
\begin{eqnarray}
{\bf C} (z) = C_u {\bf \hat x}&=&(1+\gamma){\bf \hat x}\nonumber\\
R(T)&=&1+\rho(T)\nonumber\\
\epsilon&=&1+\delta\;,
\label{expanding near critical point}
\end{eqnarray}
where $\gamma$, $\rho$ and $\delta$ are each $\ll1$. We also express the field deviation $\delta$ in terms of $\rho$ and a new effective field $h$
\begin{eqnarray}
\delta=\frac{15}{8}(h+\rho)\;.
\label{new field h}
\end{eqnarray}
We will see that this results in the critical point being located at $\rho=h=0$ and a first order low tilt--high tilt phase boundary for $\rho<0$ located along $h=0$. Inserting the change of variables of Eqs.~(\ref{expanding near critical point}) and (\ref{new field h}) into the free energy density of Eq.~(\ref{F/V rescaled}) results in the following uniform free energy near the critical point:
\begin{eqnarray}
\frac{F_u}{V}=-\frac{27u^3}{200 v^2}\bigg[ \frac{\rho}{2}\gamma^2 + \frac{1}{3}\gamma^4-h\gamma + \mathcal{O}(\gamma^5)\bigg]\;.
\label{F_u}
\end{eqnarray}
Again, since $u<0$, the overall coefficient is positive. The expression inside the brackets is the standard dimensionless Ising model free energy, which describes the Ising paramagnetic--ferromagnetic transition or the gas--liquid transition \cite{ChaikinLubensky}. In this case it describes the low tilt ($\gamma<0$)--high tilt ($\gamma>0$) Sm-$C$ transition. The critical point, where $\gamma=0$, is located at $\rho=h=0$ and the first order low tilt--high tilt phase boundary exists for $\rho<0$ along $h=0$. Note that the $\gamma^4$ term is required to stabilize the system when $\rho\leq0$. Near the critical point the equation of state is found by minimizing $F_u$ with respect to $\gamma$, i.e.,
\begin{eqnarray}
h(\gamma,\rho)=\rho\gamma+\frac{4}{3}\gamma^3\;.
\label{gamma min}
\end{eqnarray}

Next we analyze the stability of the uniform state at and near the $\rho=h=0$ critical point. We again consider the ansatz of Eq.~(\ref{ansatz 2}), but near the critical point, i.e., 
\begin{eqnarray}
C_x(Z)&=&C_{u_c}(1+\gamma)+s\cos(\alpha)\sin\bigg(\frac{qZ}{q_0}\bigg) =C_{u_c}(1+\gamma)+s\sigma\sin\bigg(\frac{qZ}{q_0}\bigg) \;,\nonumber\\
C_y(Z)&=&s\sin(\alpha)\cos\bigg(\frac{qZ}{q_0}\bigg)=s\sqrt{1-\sigma^2}\cos\bigg(\frac{qZ}{q_0}\bigg)\;,
\label{ansatz 2 near critical point}
\end{eqnarray}
and follow the same steps leading to Eq.~(\ref{q min}) and (\ref{sigma min}), to minimize the free energy with respect to $q$ and $\sigma$. Then we insert $\sigma_{\min}$ and $q_{\min}$, given by Eqs. (\ref{q min}) and (\ref{sigma min}) (but with $C_u$ replaced by $C_{u_c}(1+\gamma)$) back into the free energy. This free energy is then expanded in powers of $\gamma$ and $s$ to yield
\begin{eqnarray}
F=F_u-\frac{27u^3}{200 v^2}\bigg[ A(\gamma,\rho)s^2+B(\gamma)s^4+\mathcal{O}(s^6)\bigg]\;.
\label{F near critical point}
\end{eqnarray}
The quartic coefficient $B(\gamma)$ is, 
\begin{eqnarray}
B(\gamma)=\frac{1}{240m^4}\bigg[ 4+18m^2+8m^4 + (36m^4+18m^2+10)\gamma +
(18m^4-135m^2-75)\gamma^2 \bigg]        +        \mathcal{O}(\gamma^3)\;.
\label{B}
\end{eqnarray}
As discussed in Section \ref{sec:EstimatingLengthScaleRatio}, the dimensionless quantity $m$ is basically the ratio of length scales $\frac{\xi_{y_c}}{P_0}$ which grows with enantiomeric excess. Note that $B$ is positive at the critical point where $\gamma=0$. Thus, it is the coefficient $A(\gamma,\rho)$ that determines whether the uniform state is unstable to the modulated state, in which $s\neq0$. In terms of $m$ it is 
\begin{eqnarray}
A(\gamma,\rho)=\frac{1}{30m^2}\bigg[ -(m^2-1)^2 + \frac{15}{2}\rho m^2 -2(m^2-1)\gamma +(17m^2+12)\gamma^2 + \mathcal{O}(\gamma^3)\bigg] \;.
\label{A(gamma,rho)}
\end{eqnarray}
We also note that both $A$ and $B$ are even in $m$, confirming that the results are the same for opposite handedness. 

Before we use the above expression to find the uniform-modulated boundary, it is illustrative to consider the system at the critical point, where $\rho=h=0$. If the system is in the uniform state, then $\gamma=0$ at the critical point and 
\begin{eqnarray}
A(\gamma=0,\rho=0)=-\frac{(m^2-1)^2}{30m^2}<0\;.
\label{A at critical point}
\end{eqnarray}
Since $A$ is negative at the critical point for all $m \neq1$ it would initially appear that the uniform state at the critical point is always unstable to the modulated state. However, one must check to make sure that the corresponding modulated ansatz of Eq.~(\ref{ansatz 2 near critical point}) is physically reasonable. In particular $q_{\min}=2q_0 \sigma \sqrt{1-\sigma^2}$ must be real, which means that $\sigma_{\min}$ must be less than 1. The expression for $\sigma_{\min}$ is given in Eq.~(\ref{sigma min}) and involves both $C_u$ and $s$. At the critical point $C=C_{u_c}=1$, and we find $s_{\min}$ by minimizing the free energy of Eq.~(\ref{F near critical point}) with respect to $s$, i.e., 
\begin{eqnarray}
s_{\min_c}=\sqrt{\frac{-A(\gamma=0,\rho=0)}{2B(\gamma=0)}} \;.
\label{s_min_c}
\end{eqnarray}
Setting $s=s_{\min_c}$ and $C=C_{u_c}=1$ in the expression for $\sigma_{\min}$ yields a purely $m$ dependent expression for $\sigma$,
\begin{eqnarray}
\sigma(m)=\sqrt{1-(m^2-1)w(m) }\;,
\label{sigma(m)}
\end{eqnarray}
where $w(m)>0$ for all $m$:
\begin{eqnarray}
w(m)= \frac{3m^{10}+107m^8+362m^6+294m^4+118m^2+16}{2m^2(3m^{10}+68m^8+310m^6+328m^4+167m^2+24)}\;.
\label{w(m)}
\end{eqnarray}
Note that if $\lvert m\rvert <1$ then $\sigma(m)>1$, and $q_\text{min}$, as per Eq.~(\ref{q min}), is imaginary. Thus, the modulated ansatz of Eq.~(\ref{ansatz 2 near critical point}) is only physically reasonable for $\lvert m\rvert >1$, e.g., for sufficiently large enantiomeric excess. Thus, systems with $\lvert m\rvert <1$ are Type I and those with  $\lvert m\rvert >1$ are Type II. We also note that {\it at} $m=1$, $\sigma=1$ and the modulation wavevector $q_{\min}=0$, corresponding to a uniform phase. We will come back to this point when we further discuss how a system evolves between Type I and II behavior and vice versa.
If $m<1$ then $q_{\min}$ of the modulated ansatz of Eq.~(\ref{ansatz 2 near critical point}) is imaginary, then the ansatz no longer oscillates sinusoidally, but instead exhibits exponential growth or decay. While such an ansatz is not physically reasonable, it does give a hint of what sort of modulated structure may exist near the critical point when $\lvert m \rvert<1$, namely one with large scale uniform domains periodically broken up by regions of short scale (exponential) twist. Before we analyze the system for $\lvert m \rvert<1$, we find the second order phase boundary of the ``nose'' of the modulated state for $\lvert m \rvert>1$ in Type II systems.

\subsection{Mapping the Second Order Uniform--Modulated Phase Boundary for Type II Systems}
\label{sec:MappingSecondOrderUMPhaseBndyTypeII}

\begin{figure}[ht]
\includegraphics[scale=0.5]{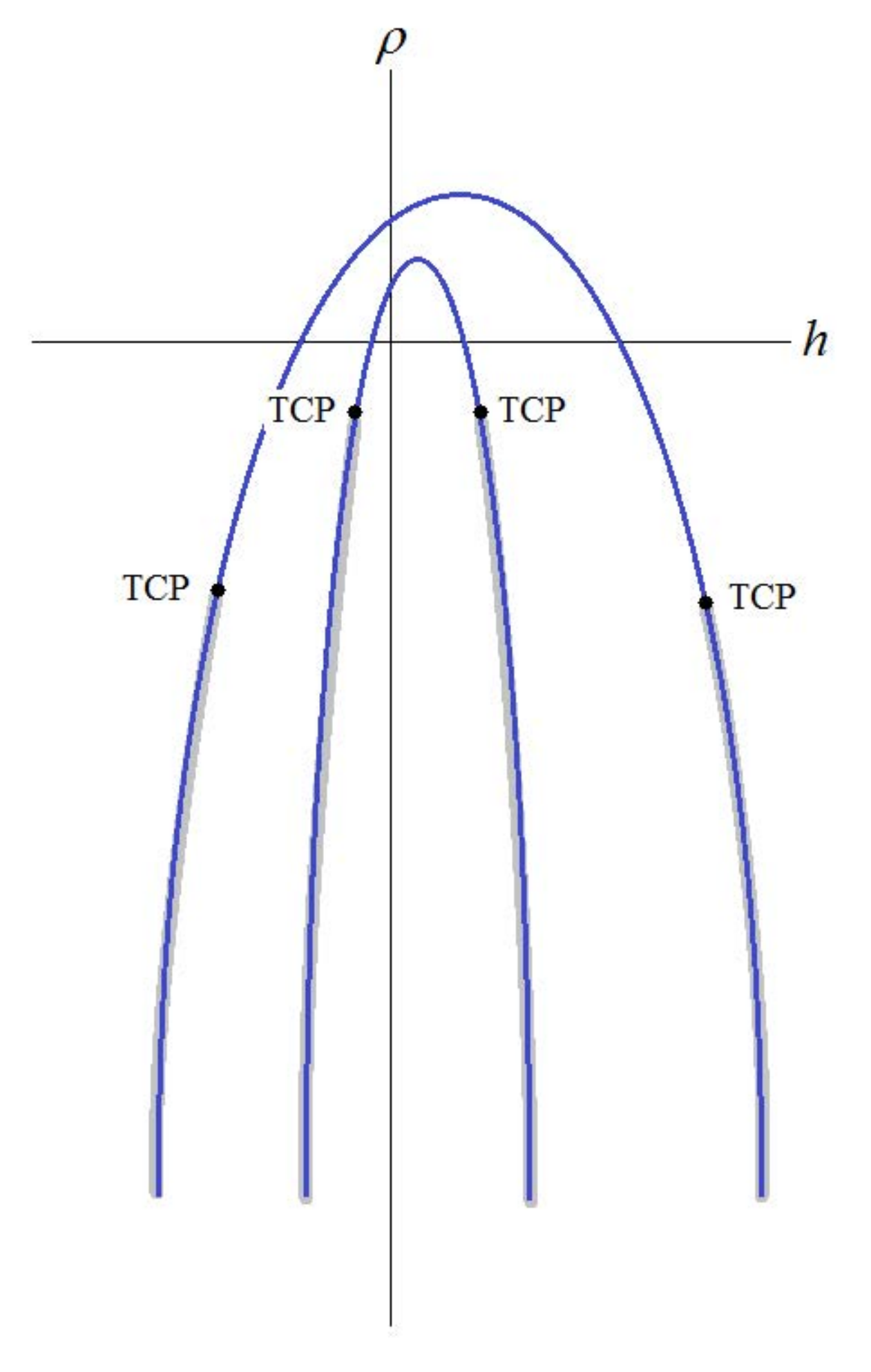}
\caption{Modulated region shrinks around unstable uniform critical point as Type I behavior is approached. Between the two tricritical points (TCP) the uniform-modulated transition is continuous. The parabolic shape of the phase boundaries, i.e., $\rho-\rho_v \propto (h-h_v)^2$, is only valid between the two TCPs. The narrower modulated region corresponds to smaller $m$, Note that for smaller $m$ the tricritical points are closer together. 
\label{shrinking modulated region}} 
\end{figure}

Next we determine the second order phase boundary in $\rho$-$h$ space for the continuous uniform-modulated (Sm-$C$--Sm-$C^*$) transition where the modulation amplitude $s_{\min}=\sqrt{\frac{-A(\gamma,\rho)}{2B(\gamma)}}$ grows continuously from zero. We consider Type II systems with enantiomeric excess just larger than the critical value, i.e., $m^2\gtrsim1$. For such systems the phase boundary is close to the critical point and the expansion in powers of $\gamma$ (Eqs.~(\ref{F near critical point}) -- (\ref{A(gamma,rho)})) is valid. The uniform modulated phase (U-M) phase boundary $\rho_{_{U-M}}(h)$is found by solving the pair of equations: $A(\gamma_{_{U-M}},\rho_{_{U-M}})=0$ and the equation of state $h=h(\gamma_{_{U-M}},\rho_{_{U-M}})$ given by  Eq.~(\ref{gamma min}). In the limit of $m^2\gtrsim1$ , the condition $A(\gamma_{_{U-M}},\rho_{_{U-M}})=0$ corresponds to
\begin{eqnarray}
58\gamma_{_{U-M}}^2-4(m^2-1)\gamma_{_{U-M}}+15\rho_{_{U-M}}=0\;.
\label{A=0 condition}
\end{eqnarray}
Solving this equation for $\gamma_{_{U-M}}$ and inserting the solution into the equation of state $h=h(\gamma_{_{U-M}},\rho_{_{U-M}})$, Eq.~(\ref{gamma min}), yields the following uniform-modulated phase boundary $\rho_{_{U-M}}(h)$
\begin{eqnarray}
\rho_{_{U-M}}(h)=\rho_v-\frac{\lambda}{(m^2-1)^4}\left(h-h_v\right)^2\;,
\label{rho_UM}
\end{eqnarray}
where $\rho_v\equiv \frac{2(m^2-1)^2}{435}$, $h_v\equiv \frac{26(m^2-1)^3}{121,945}$ give the location of the nose of the vertex of the parabola, and the constant $\lambda=\frac{307,667,235}{111,392}\approx 2,762$. We remind the reader of our sequence of rescaling and shifting of the temperature dependent parameter ($r(T)\rightarrow R(T) \rightarrow \rho(T)$) and the electric field ($E\rightarrow \epsilon \rightarrow h$). As mentioned in Section \ref{sec:EstimatingLengthScaleRatio} where we first introduced the rescaling we will outline (in Section \ref{sec:Mapping ExperimentTETheoryRepsilon}) the way in which one can convert an experimental phase diagram from $T$-$E$ space to $R(T)$-$\epsilon$ space and $\rho(T)$-$h$ space so that experimental results can be directly compared with the theoretical results, e.g., Eq.~(\ref{rho_UM}) and Fig.~\ref{shrinking modulated region}.

We emphasize that the above phase boundary is second order, i.e., only valid for continuous uniform--modulated phase transitions. Numerical analysis also reveals the existence of two tricritical points on either side of the nose, also shown in Fig.~\ref{shrinking modulated region}. Below each tricritical point the phase boundary becomes first order, corresponding to a discontinuous uniform--modulated phase transition, whereby there is a jump in the modulation amplitude $s$. Thus the above, parabolic, expression for the phase boundary is only valid {\it between} the tricritical points. As $\lvert m \rvert \rightarrow1$, these two tricritical points approach each other and seem to merge at $m=1$. In other words, $\lvert m \rvert \rightarrow1$, the second order phase boundary shrinks to zero, resulting in a phase boundary that is purely first order. 

Moreover, as $\lvert m \rvert \rightarrow1$ and the two tricritical points approach each other, the curvature of the continuous phase  boundary increases rapidly due to the $(m^2-1)^4$ in the denominator of Eq.~(\ref{rho_UM}), as shown in Fig.~\ref{shrinking modulated region}. Thus, modulated ``nose'' of the phase boundary becomes sharper and approaches a cusp-like point at the transition to Type I behavior.

As we prepare to move onto analysis of the Type I phase boundary, we recall from Section \ref{sec:ModelTypeIIBehaviorCriticalLengthScaleRatio} that the breakdown of the sinusoidally oscillating ansatz (i.e., imaginary $q$) suggests that the nature of the modulated state for $\lvert m \rvert<1$ changes to one with large scale uniform domains periodically broken up by regions of short scale (exponential) twist. This suggests that for $\lvert m \rvert<1$ we should consider a modified modulated ansatz with finite amplitude, and also anticipate a first order transition to the uniform state, which is consistent with the vanishing of the second order boundary discussed above.

We close this section by noting that location of the tricritical points analytically requires the use of an ansatz with higher wave vector harmonics\cite{Schaub and Mukamel}. This is a cumbersome process so we instead rely on the numerical analysis discussed in Section \ref{sec:NumericalAnalysis}. Lastly, in the Appendix we further analyze the modulated state just inside the nose, and show that there is no discontinuity of spatially averaged tilt within the modulated region. 

\section{Mapping the Modulated Phase Boundary for Type I Systems}
\label{sec:MappingModulatedPhaseBndyTypeIMUTriplePoint}

As discussed in the preceding section, the sinusoidally oscillating ansatz of Eq.~(\ref{ansatz 2 near critical point}) breaks down for $\lvert m \rvert<1$. Therefore we consider a different ansatz near the {\it stable} uniform critical point where ${\bf C}_{u_c}(z)={\bf \hat x}$:
\begin{eqnarray}
C_x(z)=1+\lvert\gamma\rvert\sin\phi(Z) \;,\nonumber\\
C_y(z)=\lvert\gamma\rvert\cos\phi(Z)\;,
\label{ansatz 3 near critical point}
\end{eqnarray}
where we take $\lvert\gamma\rvert$, the magnitude of the deviation of the tilt from critical value, to be independent of position. Both $\phi(Z)$ and $\lvert\gamma\rvert$ are now to be determined via minimization of the free energy. Unlike Eq.~(\ref{ansatz 2 near critical point}) we have not constrained the functional form of $\phi(z)$, which allows us to consider modulations more general than sinusoidal. Insertion of the above ansatz into the free energy Eq.~(\ref{F/V}) yields
\begin{eqnarray}
\frac{F_M}{V}=\frac{q}{2\pi q_0}\bigg(\frac{-27u^3}{200 v^2} \bigg) \int_0^{\frac{2\pi q_0}{q}} dZ \bigg[ -h\cos(\phi)\lvert\gamma\rvert+ \left(\frac{\rho}{2} +  \frac{4}{15}\sin(\phi)\right)\lvert\gamma\rvert^2 +  \frac{4}{15}\cos(\phi)\sin^2(\phi)\lvert\gamma\rvert^3 + \nonumber\\ \left(\frac{2}{5}\cos^2(\phi) - \frac{1}{15}\right)\lvert\gamma\rvert^4 
+ \frac{m^2}{15}\bigg((\partial_Z \phi)^2 -2(\partial_Z \phi)\bigg)\lvert\gamma\rvert^2 +  \mathcal{O}(\lvert\gamma\rvert^5)\bigg]\;.
\label{F_unwind near critical point1}
\end{eqnarray}
where $q_0=2\pi/P_0$ is the wave vector of the perfectly helical modulation with pitch $P_0$, and $q$ is the wave vector of the actual modulation, and is still to be determined. For a uniform state $\partial_Z \phi= 0$ and the energy is minimized by $\phi=0$ for $h>0$ and $\phi=\pi$ for $h<0$, corresponding to the high and low tilt states. In contrast to Section \ref{sec:EstimatingLengthScaleRatio} these states are distinguished by the value of the angle ($\phi=0$ for high tilt and $\phi=\pi$ for low tilt) and not by the sign of $\lvert\gamma\rvert\geq 0$ which is positive by definition. The corresponding energy of the uniform state for either the $\phi=0$ or $\phi=\pi$ states is
\begin{eqnarray}
\frac{F_{_U}}{V}=-\frac{27u^3}{200 v^2}\bigg[ \frac{\rho}{2}\lvert \gamma_{_U} \rvert^2 + \frac{1}{3}\lvert \gamma_{_U} \rvert^4-\lvert h\rvert \lvert \gamma_{_U} \rvert )\bigg]\;,
\label{uniform energy}
\end{eqnarray}
which when minimized with respect to $\lvert \gamma_{_U} \rvert$ agrees with the uniform equation of state Eq.~(\ref{gamma min}), as it should.

For the modulated state, minimizing with respect to $\phi(Z)$, and employing the Beltrami identity gives:
\begin{eqnarray}
\frac{d\phi}{dZ}=\frac{2}{m}\bigg[ W -\frac{15h}{4\lvert \gamma_{_M} \rvert}\cos(\phi) +\sin^2(\phi)\left( 1- \lvert \gamma_{_M} \rvert \cos(\phi) -\frac{3\lvert \gamma_{_M} \rvert ^2}{2 }\right) \bigg]^\frac{1}{2}\;,
\label{phi_z}
\end{eqnarray}
where $W$ is a dimensionless constant of integration that (along with $\lvert\gamma_{_M}\rvert$) is determined by further minimization. In Eq.~(\ref{phi_z}), and the analysis to follow, we take $m>0$ corresponding to positive enantiomeric excess. We note that carrying out the analysis with $m<0$ (corresponding to enantiomeric excess of opposite handedness) results in the same phase boundary, i.e., the results are independent of the handedness of the enantiomeric excess, as they should be. The above equation can be integrated to obtain $Z(\phi)$ and thus the modulated structure $\phi(Z)$. 
\begin{eqnarray}
Z=\frac{m}{2}\int_0^{\phi(z)}d\phi^\prime\bigg[ W -\frac{15h}{4\lvert \gamma_{_M} \rvert}\cos(\phi^\prime) +\sin^2(\phi^\prime)\left( 1- \lvert \gamma_{_M} \rvert \cos(\phi^\prime) -\frac{3\lvert \gamma_{_M} \rvert ^2}{2 }\right) \bigg]^{-\frac{1}{2}}\;,
\label{z(phi)}
\end{eqnarray}
where we remind the reader that the actual, unrescaled position $z=Z/q_0=Z P_0/ 2 \pi$.

Inserting the above expression for $\partial_Z \phi$ back into the free energy given by Eq.~(\ref{F_unwind near critical point1}), and by minimizing with respect to $W$ gives 
\begin{eqnarray}
m\pi=I(W_\text{min},h,\lvert \gamma_{_M} \rvert)\equiv\int^{2\pi}_0 d\phi\bigg[ W_\text{min} -\frac{15h}{4\lvert \gamma_{_M} \rvert}\cos(\phi) +\sin^2(\phi)\left( 1- \lvert \gamma_{_M} \rvert \cos(\phi) -\frac{3\lvert \gamma_{_M} \rvert ^2}{2 }\right) \bigg]^\frac{1}{2}\;.
\label{min W ito I}
\end{eqnarray}
Using Eqs.~(\ref{phi_z}) and (\ref{min W ito I}), the free energy density of Eq.~(\ref{F_unwind near critical point1}) can be simplified to
\begin{eqnarray}
\frac{F_M}{V}=\frac{-27u^3}{200 v^2} \bigg[ \frac{\rho}{2}\lvert \gamma_{_M} \rvert^2 + \frac{1}{3}\lvert \gamma_{_M} \rvert^4 - \frac{4}{15}W_\text{min} \lvert \gamma_{_M} \rvert^2\bigg]\;.
\label{F_unwind near critical point2}
\end{eqnarray}
which when minimized with respect to $\lvert\gamma_{_M}\rvert$ gives:
\begin{eqnarray}
\rho\lvert\gamma_{_M}\rvert+\frac{4}{3}\lvert\gamma_{_M}\rvert^3-\frac{8}{15} W_\text{min} \lvert\gamma_{_M}\rvert - \frac{4}{15}\frac{dW_\text{min}}{d\lvert \gamma_{_M} \rvert} \lvert \gamma_{_M} \rvert^2=0\;,
\label{mod gamma abs min}
\end{eqnarray}
The coupled equations (\ref{min W ito I}) and (\ref{mod gamma abs min}) could in principle be solved to obtain $W_\text{min}(h,\rho,m)$ and $\gamma_M(h,\rho,m)$ for the modulated state, which could in turn be used to find the modulation period in terms of the system parameters $h,\rho,m$.
\begin{eqnarray}
P(h,\rho,m)=\frac{mP_0}{4\pi}\int^{2\pi}_0 d\phi\bigg[ W_\text{min} -\frac{15h}{4\lvert \gamma_{_M} \rvert}\cos(\phi) +\sin^2(\phi)\left( 1- \lvert \gamma_{_M} \rvert \cos(\phi) -\frac{3\lvert \gamma_{_M} \rvert ^2}{2 }\right) \bigg]^{-\frac{1}{2}}\;,
\label{P}
\end{eqnarray}
So far, the above analysis is structurally similar to the standard unwinding analysis of\cite{Schaub and Mukamel} (which we also employed in Section \ref{sec:UniformModulatedPhaseBoundaryFarBelow}). By ``unwinding'' we refer to the divergence of the modulation period at the transition. The next step in this type of analysis would be to solve for $W_\text{min}$ at the unwinding transition, which we denote $W_U$. This is most easily done using Eq.~(\ref{phi_z}). For $h>0$, both $\phi \rightarrow 0$ and $\partial_Z \phi  \rightarrow 0$ as the modulated state unwinds, which means $W_{U_{h>0}}\rightarrow \frac{15h}{4\lvert \gamma_{_M} \rvert}$ at the unwinding transition. Conversely, for $h<0$, $\phi \rightarrow \pi$ and $\partial_Z \phi  \rightarrow 0$ and $W_{U_{h<0}}\rightarrow -\frac{15h}{4\lvert \gamma_{_M} \rvert}$. One then simply inserts the expression for $W_{U_{h>0}}$ (or $W_{U_{h<0}}$) into Eq.~(\ref{min W ito I}) and Eq.~(\ref{mod gamma abs min}), and solves for the corresponding $h_{_{UM}}(\rho,m)$ at which the modulated system unwinds. We note that insertion of $W_{U_{h>0}}$ (or $W_{U_{h<0}}$) into Eq.~(\ref{F_unwind near critical point2}) yields the uniform free energy Eq.~(\ref{uniform energy}), as it must since the energy of the modulated state must approach that of the uniform state as the modulation period diverges, and $\partial_Z\phi\rightarrow0$ throughout the system.

However, we will see that this standard unwinding analysis does not work in this case because the free energy of the modulated state becomes smaller than that of the uniform state while the period is still finite. In other words, the continuous unwinding transition is preempted by a first order transition between the modulated and uniform states. To find the corresponding phase boundary $\rho_{1_{st}}(h,m)$ one must instead solve the coupled Eq.~(\ref{min W ito I}) and Eq.~(\ref{mod gamma abs min}), along with the the condition $F_{_U}=F_{_M}$, where $F_{_U}$ and $F_{_M}$ are the uniform and modulated free energies given by Eqs.~(\ref{uniform energy}) and (\ref{F_unwind near critical point2}).

\subsection{Locating the Modulated-Uniform Triple Point}
\label{sec:LocatingModUniformTriplePoint}

\begin{figure}[ht]
\includegraphics[scale=0.5]{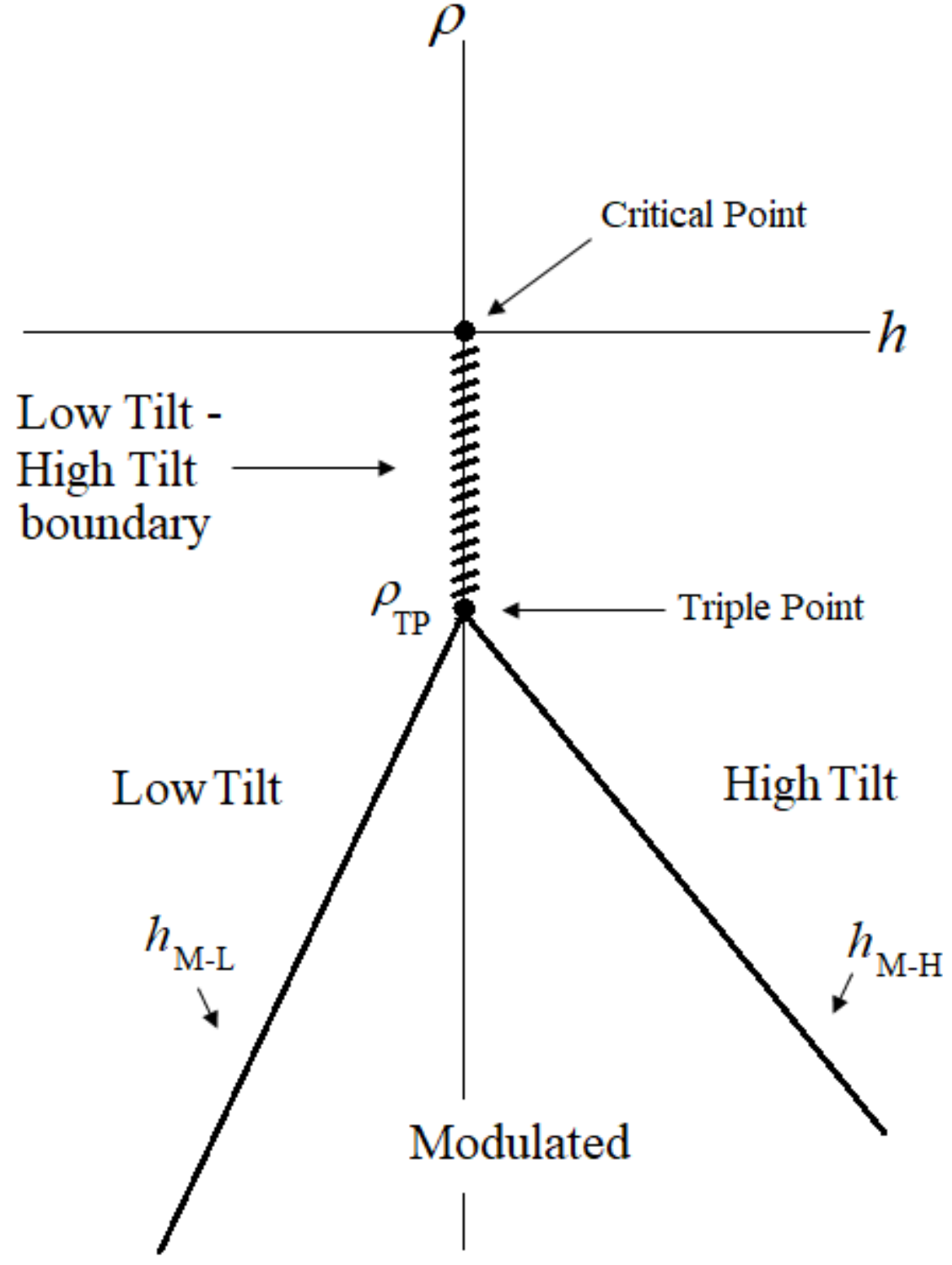}
\caption{The Type I phase diagram near the high tilt -- low tilt - modulated triple point. Note the different slopes of the phase boundary between the low tilt - modulated states and the phase boundary between the high tilt - modulated states. 
\label{Phase Boundary Near Triple Point}}
\end{figure}

We begin by locating the phase boundary at $h=0$, i.e., we find $\rho_{1_{st}}(h=0,m)$, and we will see that $\rho_{1_{st}}(h=0,m)<0$. Since the first order uniform low tilt--high tilt phase boundary exists for $h=0$\cite{h not zero boundary}  and $\rho<0$, the fact that $\rho_{1_{st}}(h=0,m)<0$ means that the two first order uniform--modulated phase boundaries must meet at a triple point, as shown in Fig.~\ref{Phase Boundary Near Triple Point}. 
Thus we denote $\rho_{1_{st}}(h=0,m)$ as $\rho_{_{TP}}(m)$. The process of finding $\rho_{_{TP}}(m)$ is somewhat cumbersome and we relegate the details to the Appendix and instead quote the result:
\begin{eqnarray}
\rho_{_{TP}}(m)=\frac{8\pi}{19}\left(m -m_1\right)\;,
\label{rhoTPm}
\end{eqnarray}
where $m_1\equiv4/\pi$. In the Appendix we show that the above results is only valid for $m\leq m_1$, i.e., $\rho_{_{TP}}\leq0$, implying that for a Type I system the triple point will always lie below the critical point as shown in Fig.~\ref{Phase Boundary Near Triple Point}. As $m\rightarrow m_1 \equiv 4/\pi$, the triple point approaches the critical point, ultimately merging at $m=m_1$. Thus, the system is Type I for $m<m_1$. We remind the reader that in Section \ref{sec:DeterminationUniformModulatedPhaseBndyTypeII} we showed that the system is Type II for $m>m_2\equiv 1$. Of course, the system must be either Type I or Type II, so there must be a single value, $m_c$ that delineates Type I and Type II behavior. Our analysis yields $m_1=4/\pi \neq m_2 =1$. However, would be quite surprising if our analysis yielded $m_1=m_2=m_c$, given that each regime required a fundamentally different ansatz. Nonetheless, it is reassuring that $m_1$ and $m_2$ are relatively close, and we would expect that the true $m_c$ is near $m_1$ and $m_2$, i.e., $m_c\approx1$. Indeed, the numerical analysis of Section \ref{sec:NumericalAnalysis} predicts a critical value $m_c\approx m_1=m_1=4/\pi$. 

In Section \ref{sec:CrossoverBetweenTypeITypeII} we will further discuss the crossover between Type I and Type II behavior, but first  we locate the first order phase boundaries that meet at the triple point. The boundary between the low ($L$) and high ($H$) tilt uniform states has already been found and lies along $h=0$ for $\rho<0$. The other two boundaries separate the modulated ($M$) and uniform high ($H$) tilt phases, and the modulated ($M$) and uniform low ($L$) tilt phases. Notationally we refer to the three phase boundaries as $L-H$, $M-H$, and $M-L$. The $M-H$ and $M-L$ phase boundaries are found by expanding near the triple point, i.e., for $\rho \lesssim \rho_{_{TP}}$. We again relegate the details of the expansion to the Appendix and move directly to the phase boundaries: 
\begin{eqnarray}
\rho_{_{M-H/M-L}}(h,m)=\rho_{_{TP}}(m) \mp \frac{19}{24}\frac{1}{\sqrt{W_{_{TP}}(m)}}h\;,
\label{approx first order bndys}
\end{eqnarray}
with $W_{_{TP}}(m)$ is given by Eq.~(\ref{WTP}). The $M-H$ expression is valid for $h>0$ so the $M-H$ boundary has negative slope, whereas the $M-L$ expression is valid for $h<0$ so the $M-L$ boundary has positive slope. In the Appendix, we show that higher order corrections to the expressions for $\rho_{_{M-H/M-L}}$ imply that the $M-L$ boundary is steeper than the $M-H$ boundary, as shown in Fig.~\ref{Phase Boundary Near Triple Point}. Recalling that $W_{_{TP}}\rightarrow0$ as $m\rightarrow m_1$, we see that the boundaries become increasingly steep, as the transition to Type II behavior is approached.

\subsection{Structure of the Modulated Phase Near the Triple Point in Type I Systems}
\label{sec:StructureModulatedPhaseNearTripleTypeI}

The pitch of the modulated phase at the triple point can be found by setting $h=0$, $W=W_{_{TP}}$ and $\gamma_{_M}= \lvert \gamma_{_{TP}} \rvert$ in Eq.~(\ref{P}). Expanding for small $W_{_{TP}}$ gives;
\begin{eqnarray}
P_{_{TP}}\approx -\frac{mP_0}{2\pi}\ln(W_{_{TP}})\;.
\label{P_TP}
\end{eqnarray}
which, as one would expect, diverges as $m\rightarrow m_1$. Thus, at $m=m_1$ where the Type I triple point and critical point merge, the modulation wavevector $q$ vanishes, corresponding to a uniform system. We recall that for $m=m_2$, where the two Type II tricritical points merged, the modulation wavevector $q$ also vanishes. As discussed above, there must be a single $m_c$ at which the system crosses over from one Type to the other, and it is reasonable to assume that the wavevector $q$ vanishes as this $m_c$ is approached from either the Type I or Type II side.

\begin{figure}[ht]
\includegraphics[scale=0.5]{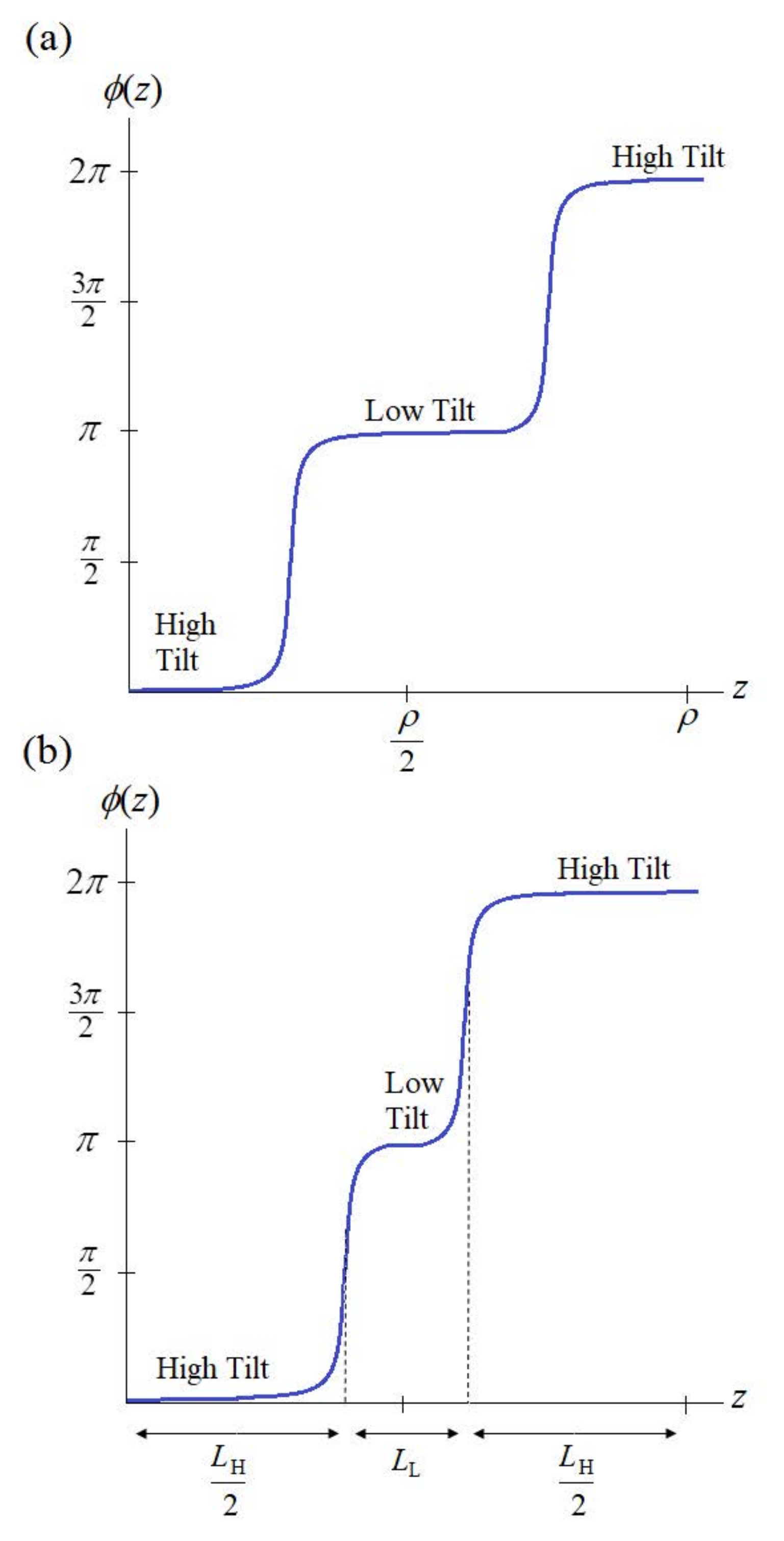}
\caption{(a)$\phi(z)$ of the modulated phase at the triple point, i.e., $\rho=\rho_{_{TP}}$ and $h=0$. The low and high tilt domains are the same size ($\approx P/2$). (b) $\phi(z)$ of the modulated phase below the triple point, $\rho\lesssim \rho_{_{TP}}$ and $h\gtrsim0$. The high tilt domains are larger than the low tilt domains.
\label{phi(z)}}
\end{figure}

\begin{figure}[ht]
\includegraphics[scale=0.5]{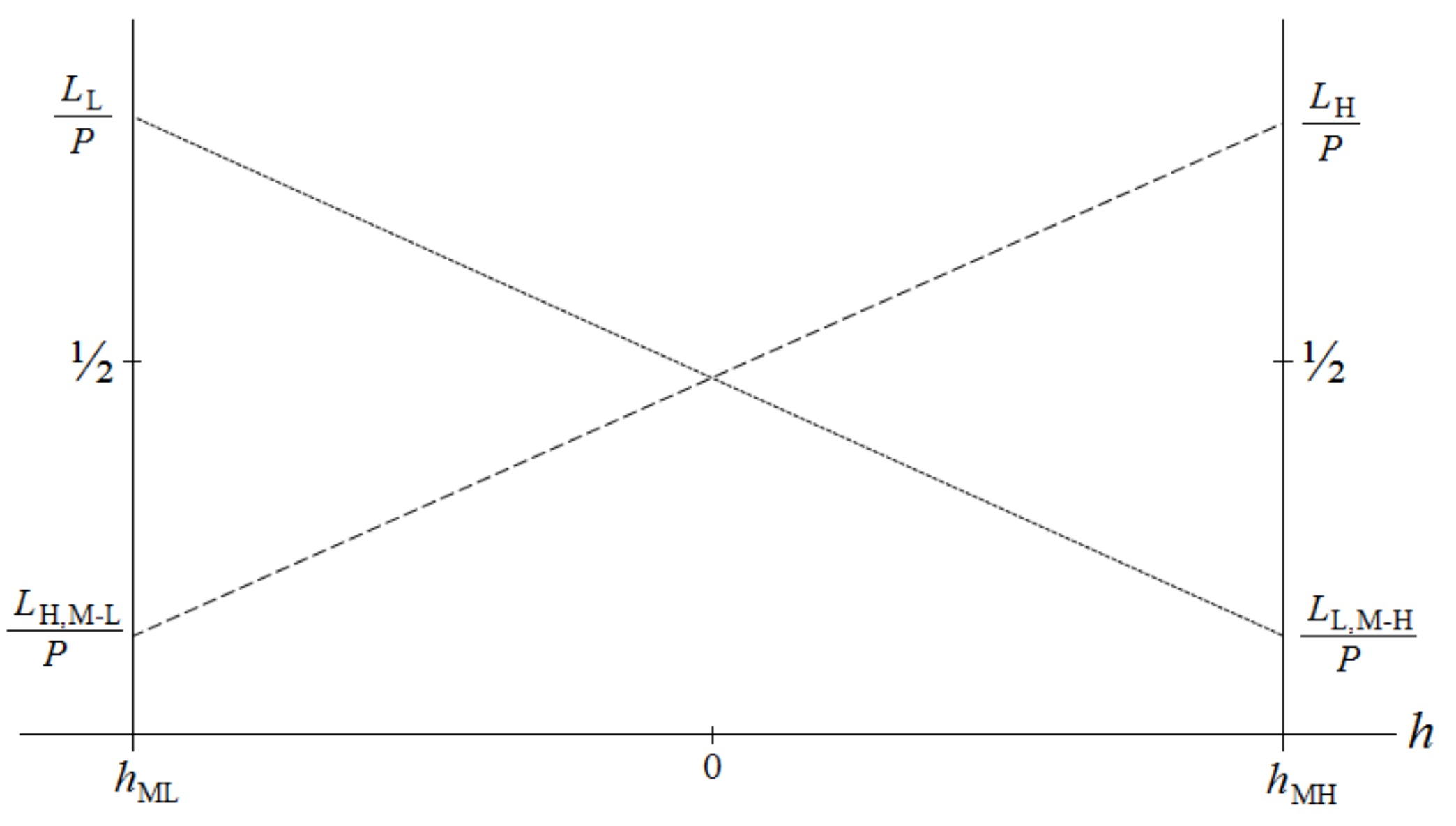}
\caption{Variation of fractional low and high tilt domain sizes, $L_L/P$ and $L_H/P$, as well as $L_{M-L}/P$ and $L_{M-H}/P$.
\label{PL vs h}}
\end{figure}

The structure $\phi(z)$ of the modulated phase at the triple point can be found using Eq.~(\ref{z(phi)}). Doing so at the triple point results in the $\phi(z)$ shown in Fig.~\ref{phi(z)}(a). This corresponds to equally long high and low tilt ($\approx \frac{P_{_{TP}}}{2}$) domains separated by narrow domain walls, as shown schematically in Fig.~\ref{Domain Walls}.
For $\rho \lesssim \rho_{_{TP}}$ the variation of $L_L$ and $L_H$ with $h$ is shown schematically in Fig.~\ref{PL vs h}.
As $\rho$ is lowered from $ \rho_{_{TP}}$, while keeping $h=0$, the modulation period becomes shorter. The relative length of the high and low tilt domains remains equal. For $h>0$, the high-tilt domains become longer than the low-tilt domains, and vice versa for $h<0$. As shown in Fig.~\ref{phi(z)}(b), one can define a low-tilt domain length:
\begin{eqnarray}
L_L=\frac{mP_0}{4\pi}\int_{\pi/2}^{3\pi/2}d\phi^\prime\bigg[ W -\frac{15h}{4\lvert \gamma_{_M} \rvert}\cos(\phi^\prime) +\sin^2(\phi^\prime)\left( 1- \lvert \gamma_{_M} \rvert \cos(\phi^\prime) -\frac{3\lvert \gamma_{_M} \rvert ^2}{2 }\right) \bigg]^{-\frac{1}{2}}\;,
\label{L_L}
\end{eqnarray}
which depends on $\rho$ via $W$ and $\lvert \gamma_{_M} \rvert$, each of which would be found by solving Eqs.~(\ref{min W ito I})--(\ref{mod gamma abs min}). The corresponding high-tilt domain length is simply $L_H=P-L_L$, and for $h=0$, $L_L=L_H \approx P/2$. While one can use the above equation to find $L_L$ as a function of $\rho$, $h$, and $m$, it is rather tedious and not particularly illuminating. In particular, since the $M-H$ and $M-L$ transitions are first order, $L_L$ and $L_H$ will remain finite at  the $M-H$ and $M-L$ boundaries. 

As discussed in the Introduction the transition to the modulated phase can be thought of as occurring via the nucleation of a periodic array of domain walls, as shown schematically in Fig.~\ref{Domain Walls}. For example, as $h$ is lowered through $h_{M-H}$ the system transitions to the modulated phase from the high-tilt phase, where there is a nucleation of a periodic array of low-tilt domains, each of length $L_{L,M-H}$. This is analogous to the transition to the modulated Abrikosov flux lattice in Type II from the superconducting phase. This Abrikosov flux lattice can be thought of as a superconducting phase containing a periodic array of normal domains or defects. As $h$ continues to be lowered, the density ($L_L/P$) of the low-tilt domains grows, and eventually the system transitions to the low-tilt phase at $h_{M-L}$. Similarly the transition from the low-tilt phase to the modulated phase at $h_{M-L}$ occurs via the nucleation of an array of periodic array of high-tilt domains of length $L_{H,M-L}$. 

\section{Crossover between Type I and Type II behavior}
\label{sec:CrossoverBetweenTypeITypeII}

Now that we have obtained separate $\rho(T)$-$h$ phase diagrams, Figs.~\ref{Phase Boundary Near Triple Point} and \ref{shrinking modulated region}, near the modulated nose for Type I and Type II respectively, we discuss the cross over from one Type to the other. As discussed in Section \ref{sec:MappingModulatedPhaseBndyTypeIMUTriplePoint} our analysis for each Type of system employs a different ansatz. We show that the system will be Type I for $m<m_1\equiv 4/\pi$ and Type II for $m>m_2\equiv 1$. Of course, the system must be either Type I or Type II, so there must be a single value, $m_c$ that delineates Type I and Type II behavior. While our $m_1\neq m_2$, it is reassuring that $m_1$ and $m_2$ are relatively close, and we would expect that the true $m_c$ is near $m_1$ and $m_2$, i.e., $m_c\approx1$. Indeed, the numerical analysis of Section \ref{sec:NumericalAnalysis} predicts a critical value $m_c\approx m_1=m_1=4/\pi$. 

Based on the evolution with $m$ of the phase diagram for each Type, we can can infer how the crossover from one Type to the other occurs. Consider a Type I system, i.e., with $m<m_c$, with phase diagram shown in Fig.~\ref{Phase Boundary Near Triple Point}. At each of the two uniform--modulated first order phase boundaries, the modulation wave vector $q$ changes discontinuously (zero in the uniform phase, finite in the modulated phase). There will also be a jump in $\lvert \gamma \rvert$, and thus the tilt. These two phase boundaries (and the uniform low--uniform high tilt boundary) meet at a triple point. At the triple point, $q_{TP}$ is finite, and will jump discontinuously to zero as one crosses the triple point in transitioning to the uniform state. If one crosses the triple point to the right of the uniform low--uniform high tilt boundary, the system will jump to the high tilt state, whereas crossing to the other side will take the system to the low tilt state. Either way, there is jump in $\lvert \gamma \rvert$. 

As $m\rightarrow m_c$, the triple point approaches and merges with the critical point into what we term a  critical triple point located at $\rho=0$, $h=0$ and $m=m_c$. As discussed following Eq.~(\ref{P_TP}), the modulation period diverges at the triple point, i.e., $q_{TP}\rightarrow 0$ as $m\rightarrow m_c$. Thus, at $m=m_c$, the transition, via the critical triple point, from the modulated to uniform phase will be accompanied by a {\it continuous} vanishing of $q$. Moreover, at the critical point the discontinuity between the uniform low and high tilt states vanishes. This means that at the critical triple point, the uniform--modulated transition is continuous in {\it both} wavevector $q$ and tilt $\lvert \gamma \rvert$. In other words, at the critical triple point, the three phases (modulated, uniform low tilt, uniform high tilt) are both energetically {\it and} symmetrically equivalent. Of course, this is only true {\it at} the critical triple point. On either side of it is a first order modulated--uniform phase boundary where both wavevector and tilt jump discontinuously. Thus, the critical triple point can be thought of as a second order phase boundary of infinitesimal (point-like) extent.

As $m$ exceeds $m_c$ the system becomes Type II. Now the critical triple point splits into two tricritical points, in between which there is a second order modulated--uniform phase boundary as shown in Fig.~\ref{shrinking modulated region}. In other words the infinitesimal (point-like) second order boundary at $m=m_c$ now grows in extent as $m$ exceeds $m_c$. In addition, as $m$ exceeds $m_c$ the cusp-like modulated--uniform phase boundary becomes parabolic, with continuously decreasing curvature. On this second order phase boundary it is the amplitude of the finite $q$ modulation that grows continuously as one crosses to the modulated phase. This is sometimes referred to as an ``instability'' type transition, as opposed to the ``unwinding'' type transition discussed in Section \ref{sec:UniformModulatedPhaseBoundaryFarBelow} that occurs via the continuous growth of the modulation wave vector. Thus, the critical triple point located at $\rho=0$, $h=0$ and $m=m_c$ can be thought of as a point where the transition is simultaneously instability type {\it and} unwinding type. 

\begin{figure}[ht]
\includegraphics[scale=0.4]{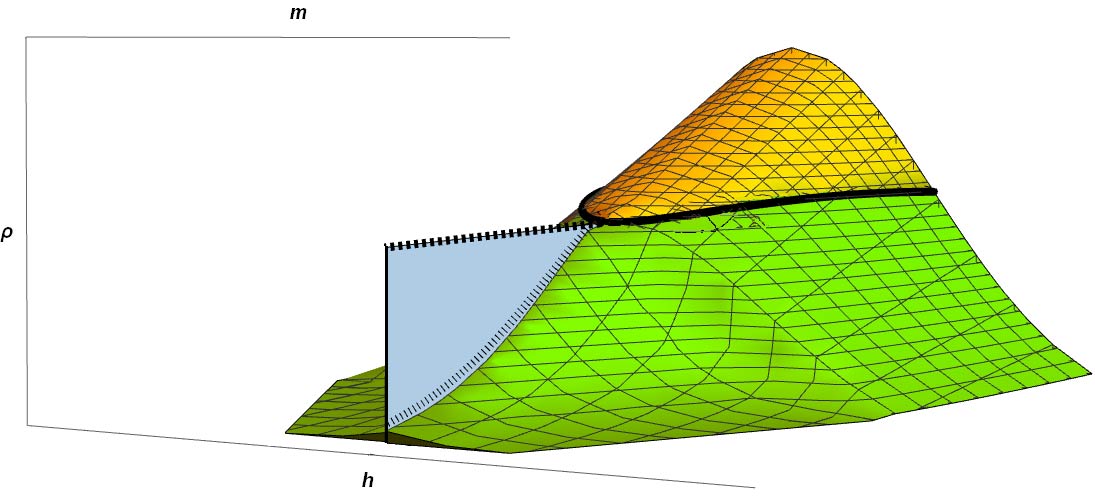}
\caption{The phase diagram in $\rho$-$h$-$m$ space near the critical triple point. The modulated phase lies below the green and gold surface, while the uniform phase lies above the surface. The green and gold surfaces correspond to first order and continuous uniform-modulated phase boundaries respectively. The gold surface is bounded by a line of tricritical points. The vertical $h=0$ plane separates the low and high tilt uniform phases, and crossing the plane corresponds to the first order uniform low--uniform high phase transition. The dotted line at the top of the vertical plane is a line of uniform low--uniform high critical points, while the dashed line at the bottom of the plane is a line of uniform low--uniform high--modulated triple points. These lines, and the line of tricritical points, all meet at $\rho=0$, $h=0$, $m=m_c$ at a what we term a critical triple point. For $m<m_c$ the system is Type I and and for $m>m_c$ it is Type II. 
\label{3DPhaseDiagram}}
\end{figure}

In Fig.~\ref{3DPhaseDiagram} we show the phase diagram in $\rho$-$h$-$m$ space, which incorporates both the Type I and Type II $\rho$-$h$ phase diagrams. The modulated phase lies below the green and gold surface, while the uniform phase lies above the surface. Crossing the green region corresponds to a first order uniform--modulated transition, while crossing the gold region corresponds to a continuous uniform--modulated transition. These two regions meet at a line of tricritical points represented by the solid line bounding the yellow region. The vertical $h=0$ plane separates the low and high tilt uniform phases, and crossing the plane corresponds to the first order uniform low--uniform high phase transition. The dotted line at the top of the vertical plane is a line of uniform low--uniform high critical points, while the dashed line at the bottom of the plane is a line of uniform low--uniform high--modulated triple points. These lines meet at $\rho=0$, $h=0$, and $m=m_c$ at a critical triple point. For $m<m_c$ the system has a uniform low--uniform high phase transition and is thus Type I. For Type I systems the transition to the modulated state is first order. Taking a $\rho$-$h$ cross-section at $m<m_c$ will yield a $\rho$-$h$ phase diagram like that shown in Fig.~\ref{Phase Boundary Near Triple Point}. For $m>m_c$ there is no longer a phase transition between the uniform phases. Instead there is an intermediate modulated phase, and the system is Type II. For $m>m_c$ the uniform-modulated transition can be first order (across the green surface) or continuous (across the gold surface). These two distinct regions meet at a line of tricritical points. Taking a $\rho$-$h$ cross-section at $m>m_c$ will yield a $\rho$-$h$ phase diagram with two tricritical points, as shown in Fig.~\ref{shrinking modulated region}.

\section{Numerical Analysis of Phase Diagrams}
\label{sec:NumericalAnalysis}

The analyses of the preceding sections have relied on the use of ansatz solutions, due to the non-harmonic nature of the free energy density, Eq.~(\ref{F/V rescaled}). Next we use brute force numerics to check that the results are reasonable. We do this by minimizing the free energy density which we represent in polar form, i.e., $C_x(z)=C(Z) \cos\phi(Z)$ and $C_y(Z)=C(Z) \sin\phi(Z)$, 
\begin{eqnarray}
\frac{F}{V}=\frac{q}{2\pi q_0}\bigg(\frac{-27u^3}{200 v^2}\bigg)\int_0^{\frac{2\pi q_0}{q}} dZ\bigg[\frac{R(T)}{2}C^2&-&\frac{1}{6}C^4+\frac{1}{30}C^6 + \frac{m^2}{15}\bigg( \left(\partial_Z C \right)^2  + C^2\left(\partial_Z \phi\right)^2 -2 C^2\left(\partial_Z \phi\right)  \bigg) \nonumber\\
&-& \frac{8}{15} \epsilon C \cos(\phi)\bigg] \;,
\label{F/V rescaled polar}
\end{eqnarray}
Minimization with respect to $C(Z)$ and $\phi(Z)$ gives:
\begin{eqnarray}
R(T)C-\frac{2}{3}C^3+\frac{1}{5}C^5-\frac{8}{15}\epsilon\cos\phi+\frac{2m^2}{15}\bigg(C\left(\partial_Z \phi \right)^2-2C\left(\partial_Z \phi \right) -\partial_Z^2C \bigg)=0\;,
\label{EL for C}
\end{eqnarray}
and
\begin{eqnarray}
\frac{8}{15}\epsilon C\sin\phi-\frac{2m^2}{15}\bigg( 2C\left( \partial_Z C\right) \left( \partial_Z \phi\right) +C^2\left( \partial_Z^2 \phi\right)-2C\left( \partial_Z C \right) \bigg)=0\;.
\label{EL for phi}
\end{eqnarray}
These two second-order, coupled, nonlinear equations are then numerically solved as a boundary value problem (BVP). We use the built-in routines in Matlab \cite{Matlab}, \texttt{bvp4c} or \texttt{bvp5c} as needed. Each utilizes the Lobatto IIIa formula, an implicit Runge-Kutta formula with a continuous extension \cite{bvp4cPaper}. 
\begin{figure}[ht]
\includegraphics[scale=0.6]{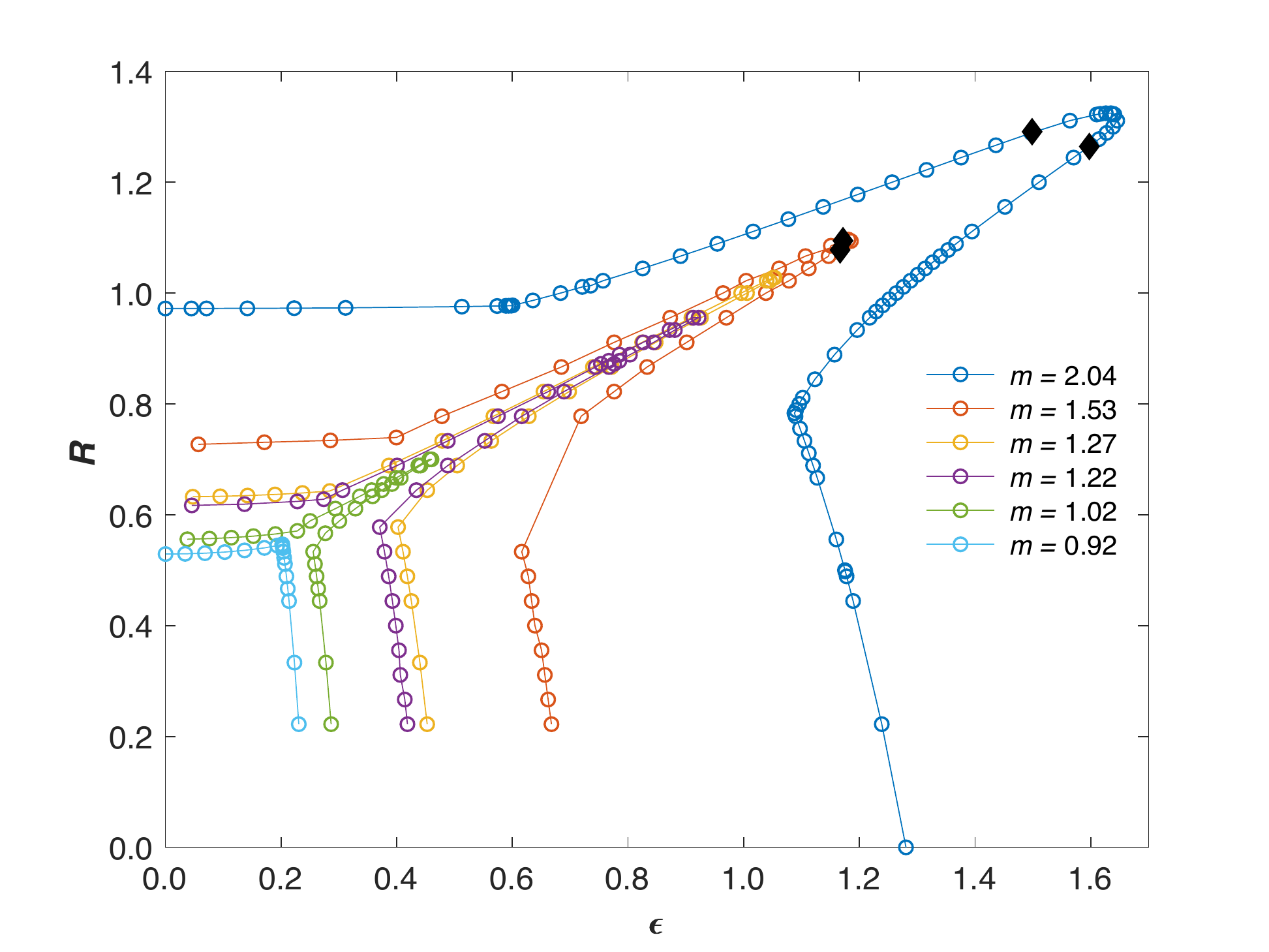}
\caption{The modulated phase boundary for systems with different $m$ values (rightmost being largest). For $m>1.27$ each phase boundary has two tricritical points, shown as black diamonds. For $m<1.27$ the tip of each ``nose" is a triple point. There is a uniform low tilt--high tilt first order boundary (not shown) that extends from each triple point to a critical point located at $R=\epsilon=1$. 
\label{Rvsepsilon}}
\end{figure}
%
\begin{figure}[ht]
\includegraphics[scale=0.6]{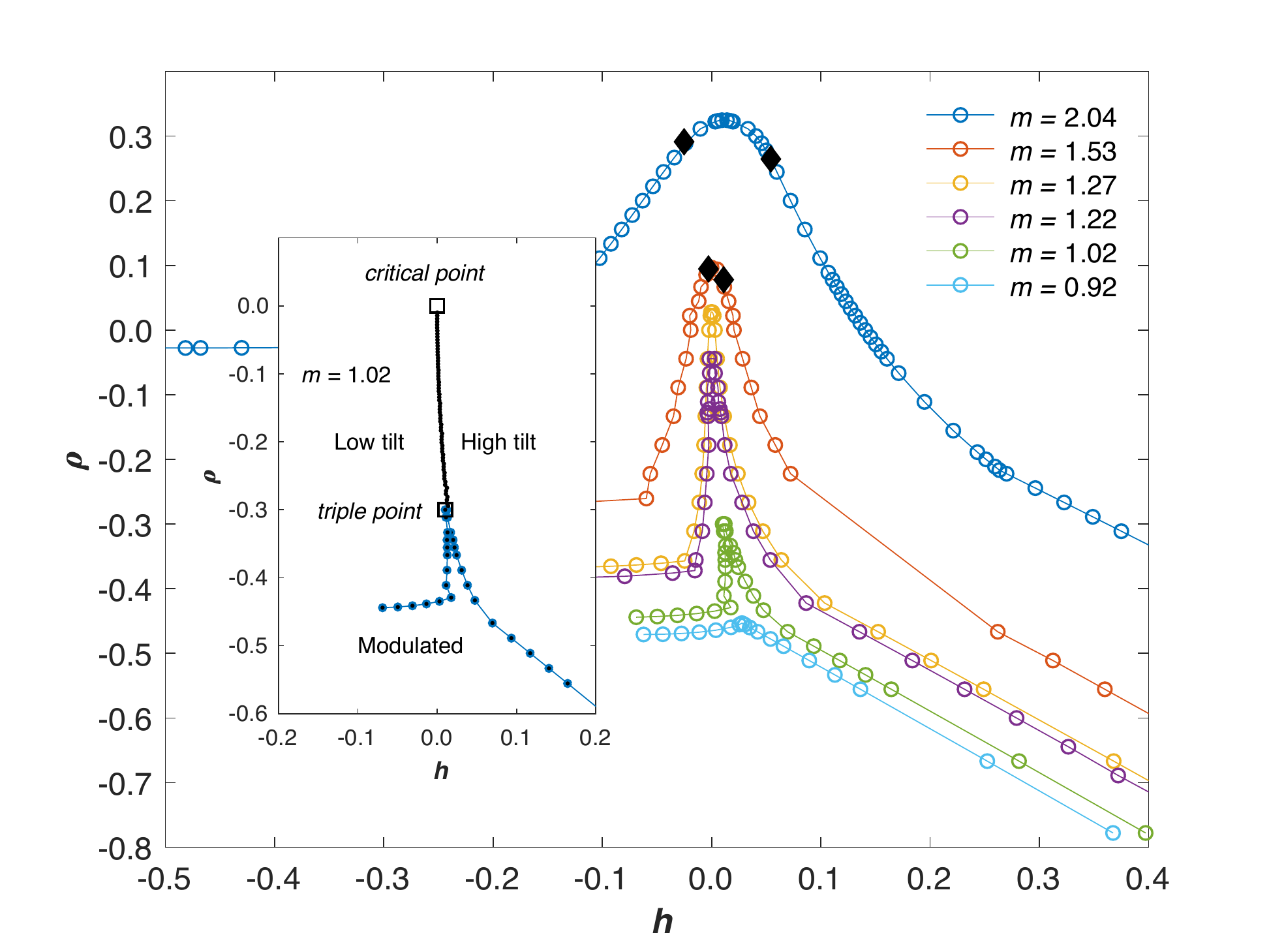}
\caption{The modulated phase boundary for systems with different $m$ values (topmost being largest). For $m>1.27$ each phase boundary has two tricritical points, shown as black diamonds. For $m<1.27$ the tip of each ``nose" is a triple point. There is a uniform low tilt--high tilt first order boundary (only shown in inset for $m=1.02$) that extends from each triple point to a critical point located at $\rho=h=0$
\label{rhovsh}}
\end{figure}

These solutions are then used to map out the phase diagram in $R$-$\epsilon$ space for a given $m$ value. Specifically, we loop through a set of different $R$ values, and for each $R$ value find the $\epsilon$ location of the phase boundary. Lowering $R$ from high to low, one will first locate the top of the modulated ``nose" region, where the phase boundary is second order, i.e., where the modulated-uniform transition is continuous. The $\epsilon(R)$ location of the boundary was found by determining where the amplitude of the modulation vanishes continuously with $\epsilon$, i.e., where the modulated solution continuously becomes uniform. Of course, the nonmonotonic ``nose" shape of the phase boundary means that two such $\epsilon$ values are found for each $R$ value. Care must be taken to compare the energies of the modulated and uniform solutions near the phase boundary. Eventually, at sufficiently low $R$ value, the energy of the uniform solution becomes smaller than the modulated solution {\it within} the modulated region. In other words the uniform phase becomes energetically preferable before the modulation vanishes, and the modulated-uniform phase transition becomes first order. The evolution from a continuous to first order transition occurs at a specific location on the phase boundary, namely at a tricritical point. Two such tricritical points are found on the ``nose," but not at the same $R$ value. For $R$ values below each tricritical point the first order boundary between the uniform and modulated phases is found by determining were the energies of the uniform and modulated solutions are equal.

Figures \ref{Rvsepsilon} and \ref{rhovsh} show phase diagrams for a variety of $m$ values in $R$-$\epsilon$ space and $\rho$-$h$ space respectively. These two phase diagrams are related by a simple shift and rotation, i.e., $\rho=R-1$ and $h=\frac{8}{15}(\epsilon-1)-\rho$. We note that the numerically obtained $\rho$-$h$ phase diagrams shown in Fig.~\ref{rhovsh} compare qualitatively well with those predicted analytically in Sections \ref{sec:MappingSecondOrderUMPhaseBndyTypeII} and \ref{sec:MappingModulatedPhaseBndyTypeIMUTriplePoint}. Also, looking at Fig.~\ref{rhovsh} one can see that the transition from Type I to Type II occurs at approximately $m_c\approx1.22$, which compares  well with the value of $m_c=4/\pi=1.27$ estimated in Section \ref{sec:MappingModulatedPhaseBndyTypeIMUTriplePoint}.

The phase diagrams in $R$-$\epsilon$ space will most resemble those in $T$-$E$ space, while those in $\rho$-$h$ space focus on the nose region and more clearly show the evolution from Type I to Type II behavior. In the next section we describe how one can convert an experimental phase diagram from $T$-$E$ space to $R(T)$-$\epsilon$ space and $\rho(T)$-$h$ space so that experimental results can be directly compared with the theoretical phase diagrams of Figs.~\ref{Rvsepsilon} and \ref{rhovsh}.

\section{Rescaling Experimental Data for Comparison with Theory}
\label{sec:Mapping ExperimentTETheoryRepsilon}

To compare experimental results with our theoretical predictions, it is necessary to convert experimental phase diagrams in $T$-$E$ space to $R(T)$-$\epsilon(E)$ space. The process of doing so differs depending on whether the phase diagram is Type I or Type II.

\subsection{Rescaling Experimental Data for Comparison with Theory: Type I}
\label{sec:RescalingExperimentalDataComparisonTypeI}

We remind the reader that $\epsilon=E/E_c$, so that for a Type I system (one with a critical point at $T_c$, $E_c$) one only need know the experimental value of $E_c$ to rescale from $E$ to $\epsilon$. The conversion from $T$ to $R$ is slightly more involved. We recall that $R(T)=r(T)/r(T_c)$ and that $r(T)=\alpha(T-T_0)$, where the constants $\alpha$ and $T_0$ are system dependent. Thus 
\begin{eqnarray}
R(T)=\frac{T-T_0}{T_c-T_0}\;,
\label{R(T)}
\end{eqnarray}
and we see that converting from $T$ to $R(T)$ requires knowing $T_0$ and $T_c$. The experimental value of $T_c$ can be obtained directly from the location of the critical point. To obtain the value of $T_0$, one can use the experimental values of $T_{_{AC}}^*$, the temperature of the zero field Sm-$A$--Sm-$C^*$ transition, and $T_{TP}$, the temperature of the triple point. Converting Eq.~(\ref{r, s E=0}) to $T_0$, $T_c$, and $T_{_{AC}}^*$ gives
\begin{eqnarray}
\frac{T_{_{AC}}^*-T_0}{T_c-T_0}=\frac{5}{12}+\frac{2m^2}{15}\;.
\label{TAC}
\end{eqnarray}
Converting Eq.~(\ref{rhoTPm}) to $T_0$, $T_c$, and $T_{_{TP}}$ gives
\begin{eqnarray}
\frac{T_{_{TP}}-T_0}{T_c-T_0}=1+\frac{8\pi}{19}\left(m-\frac{4}{\pi}\right)\;.
\label{TTP}
\end{eqnarray}
The above two equations can then be solved for $T_0$ and $m$. The $T_0$ value, along with $T_c$ can be used to convert $T$ to $R(T)$ as per Eq.~(\ref{R(T)}). This conversion along with $\epsilon=E/E_c$ then allows one to produce an $R$-$\epsilon$ phase diagram for comparison with Fig.~\ref{Rvsepsilon}. To produce a $\rho$-$h$ diagram for comparison with Fig.~\ref{rhovsh} one simply uses $R$ and $\epsilon$ along with
\begin{eqnarray}
\rho&=&R-1,\nonumber\\
h& =& \frac{8}{15}\left(\epsilon -1\right)-\rho\;.
\label{rho&h}
\end{eqnarray}

\subsection{Rescaling Experimental Data for Comparison with Theory: Type II}
\label{sec:RescalingExperimentalDataComparisonTypeII}

Unlike the phase diagram for a Type I system which has three special points (critical point, triple point and zero field Sm-$A^*$--Sm-$C^*$ transition), Type II lacks a critical point and thus only has two special points: the vertex of the parabolic modulated region and the zero field Sm-$A^*$--Sm-$C^*$ transition. The relationship between $T_0$, $T_c$, $T_{_{AC}}^*$, and $m$ is given in Eq.~(\ref{TAC}). The vertex $T_v$, $E_v$ of the parabola can be related to $T_0$, $T_c$, $E_c$ and $m$ using the expressions for $\rho_v$ and $h_v$ given after Eq.~(\ref{rho_UM}), giving:
\begin{eqnarray}
\frac{T_v-T_c}{T_c-T_0}=\frac{2\left( m^2-1\right)^2}{435}\;,
\label{Tv}
\end{eqnarray}
and
\begin{eqnarray}
\frac{8}{15}\left( \frac{E_v-E_c}{E_c}\right) = \left(\frac{T_v-T_c}{T_c-T_0}\right) + \frac{26\left( m^2-1\right)^3}{121,945}\;.
\label{Ev}
\end{eqnarray}
Together the three Eqs.~(\ref{TAC}), (\ref{Tv}), (\ref{Ev}) contain four unknowns: $T_0$, $T_c$, $E_c$ and $m$ so a fourth relationship is required. An extra relationship can be obtained using the experimental curvature at the vertex of the parabolic modulated region and the parabolic equation Eq.~(\ref{rho_UM}):
\begin{eqnarray}
\bigg\lvert \frac{d^2T}{dE^2}\bigg\rvert_v=\frac{2,946}{\left( m^2-1\right)^4}\frac{(T_c-T_0)}{E_c^2}\;.
\label{VertexCurv}
\end{eqnarray}
One can now solve Eqs.~(\ref{TAC}), (\ref{Tv}), (\ref{Ev}) and (\ref{VertexCurv}) to obtain $T_0$, $T_c$, $E_c$, and $m$, and to subsequently convert $T\rightarrow R \rightarrow \rho$ and $E \rightarrow \epsilon \rightarrow h$ as described above in the subsection for Type I.

\section{Conclusion}
\label{sec:Conclusion}

In summary, we have carried out an in-depth analysis of the electroclinic effect in ferroelectric liquid crystal systems that have a first order Smectic-$A^*$--Smectic-$C^*$ (Sm-$A^*$--Sm-$C^*$) transition, and have shown that such systems can be either Type I or Type II. In temperature--field parameter space Type I systems exhibit a macroscopically achiral (in which the Sm-$C^*$ helical superstructure is expelled) low-tilt (LT) Sm-$C$--high-tilt (HT) Sm-$C$ critical point, which terminates a LT Sm-$C$--HT Sm-$C$ first order boundary. This boundary extends to an achiral-chiral triple point at which the achiral LT Sm-$C$ and HT Sm-$C$ phases coexist along with the chiral Sm-$C^*$ phase. In Type II systems the critical point, triple point, and first order boundary are replaced by a Sm-$C^*$ region, sandwiched between LT and HT achiral Sm-$C$ phases, at low and high fields respectively. 

Whether the system is Type I or Type II is determined by the ratio of two length scales, one of which is the zero-field Sm-$C^*$ helical pitch. The other length scale depends on the size of the discontinuity (and thus the latent heat) at the zero-field first order Sm-$A^*$--Sm-$C^*$ transition. We have proposed ways in which a system could be experimentally tuned between Type I and Type II behavior, e.g., by doping a low chirality Sm-$C^*$ system with a high chirality, tight-pitch Sm-$C^*$ compound. We have also shown that this Type I vs Type II behavior is the Ising universality class analog of Type I vs Type II behavior in XY universality class systems. Specifically, the LT and HT achiral Sm-$C$ phases are analogous to normal and superconducting phases, while the 1D periodic Sm-$C^*$ superstructure is analogous to the 2D periodic Abrikosov flux lattice. 

We have made (analytically and numerically) a complete mapping of the phase boundaries in temperature--field parameter space  and show that a variety of interesting features are possible, including a multicritical point, tricritical points and a doubly reentrant Sm-$C$--Sm-$C^*$-Sm-$C$--Sm-$C^*$ phase sequence. In addition we have shown how the system crosses over between Type I and Type II behaviors.

In future, we plan to expand our model to consider thermal fluctuations about the ground states, particularly in the region of parameter space where the Type I -- Type II crossover occurs, i.e., at the critical triple point. We also plan to consider the effect electric field terms that are higher order than the linear term considered here. Preliminary analysis suggests that such terms may compete with the linear term, and may even produce a critical point in a system with a continuous zero field Sm-$A$--Sm-$C^*$ transition.

\appendix*
\section{Details of Phase Boundary Calculations}
\label{sec:Appendix}

\subsection{Showing that the Tilt is Continuous in the Modulated Phase for Type II Systems.}
\label{sec:ShowingTiltContinuousModulatedPhaseTypeII}

As shown in Section \ref{sec:DeterminationUniformModulatedPhaseBndyTypeII}, for Type II systems the uniform low tilt -- high tilt critical point is unstable to a modulated phase. We recall that this critical point terminated the uniform low tilt -- high tilt first order phase boundary, and that the tilt changed discontinuously across the boundary. We now show that the average tilt in the modulated phase remains continuous, i.e., both the critical point and tilt discontinuity are absent. Our starting point is the spatially averaged tilt, i.e, the tilt averaged over one period:
\begin{eqnarray}
\langle {\bf c} \rangle = \frac{q}{2\pi}\int_0^{\frac{2\pi}{q}} {\bf c}(z)dz=c_u {\bf \hat x}=c_{u_c}(1+\gamma){\bf \hat x}\;.
\label{average tilt}
\end{eqnarray}
Near the phase boundary given by Eq.~(\ref{rho_UM}), the system is close to the unstable uniform critical point, so as in the previous section, we work with the small deviation $\gamma$. Of particular interest is whether the 
discontinuous behavior of $\gamma$, and thus the average tilt, still exists despite the modulation. Looking at the expression for $F_u$, the free energy of the uniform state (when $s=0$), given by Eq.~(\ref{F_u}), it is useful to recall that for $\rho<0$, the $\gamma^2$ coefficient is negative and discontinuous behavior of $\gamma$ is possible. 

To determine whether this discontinuous behavior of $\gamma$ is possible in the modulated state, we must determine the $\gamma^2$ coefficient when $s\neq0$. We do so by inserting $s_{\min}=\sqrt{\frac{-A(\gamma,\rho)}{2B(\gamma)}}$ back into the free energy of Eq.~(\ref{F near critical point}). Expanding for $m^2\gtrsim1$ gives a purely $\gamma$ dependent energy within the modulated phase, to $\mathcal{O}(\gamma^3)$:
\begin{eqnarray}
F_M(\gamma)&=&F_u-\frac{27u^3}{200 v^2}\bigg[ \left( \frac{29}{30}(\rho_c-\rho)+\frac{26}{225}(m^2-1)^2\right)\gamma^2 - \left(\frac{1}{15}(\rho_c-\rho)(m^2-1) + \frac{56}{6525}(m^2-1)^3\right) \gamma \bigg] \nonumber\\
&=&-\frac{200 v^2}{27u^3}\bigg[ \left( \frac{7}{15}(\rho_c-\rho)+\frac{729}{6525}(m^2-1)^2\right)\gamma^2 -\left(h+\frac{1}{15}(\rho_c-\rho)(m^2-1) + \frac{56}{6525}(m^2-1)^3\right) \gamma )\bigg]\;.
\label{F_M gamma}
\end{eqnarray}
Remembering that within the modulated phase, $\rho<\rho_c$, we see that the $\gamma^2$ coefficient in $F_M(\gamma)$ is always positive, implying that there is no discontinuous behavior of $\gamma$ in the modulated phase.

\subsection{Locating the Triple Point in Type I Systems}
\label{sec:LocatingTriplePointTypeI}

To obtain $\rho_{1_\text{st}}(h=0,m)\equiv\rho_{_{TP}}(m)$ we begin by setting $h=0$ in Eq.~(\ref{min W ito I}) and expanding the right hand side for small $\lvert \gamma_{_M} \rvert$ and small $W$. Doing so yields a relationship between $\lvert \gamma_{_M} \rvert$, $W$ and $m$:
\begin{eqnarray}
\lvert \gamma_{_M} \rvert^2 \approx G(W,m) \equiv -\frac{3}{4}\rho_m+\frac{6}{19}\left[4\ln(2)+1-\ln(W) \right]W\;,
\label{gamma ito W}
\end{eqnarray}
where $\rho_m = \frac{8}{19}\left( \pi m-4\right)$. The condition that the modulated and uniform free energies (given by Eqs.~(\ref{uniform energy}) and (\ref{F_unwind near critical point2})) are equal implies that
\begin{eqnarray}
\frac{\rho_{_{TP}}}{2}\lvert \gamma_{_M} \rvert^2 + \frac{1}{3}\lvert \gamma_{_M} \rvert^4 - \frac{4}{15}W_{_{TP}} \lvert \gamma_{_M} \rvert^2 = \frac{\rho_{_{TP}}}{2}\lvert \gamma_{_U} \rvert^2 + \frac{1}{3}\lvert \gamma_{_U} \rvert^4\;.
\label{Energy Diff}
\end{eqnarray}
We then use Eq.~(\ref{gamma ito W}) to eliminate $\lvert \gamma_{_M} \rvert$ and reexpress Eq.~(\ref{mod gamma abs min}) and Eq.~(\ref{Energy Diff}) purely in terms of $m$ and $W_{_{TP}}$ (the value of $W$ at the triple point):
\begin{eqnarray}
\rho_{_{TP}}+\frac{4}{3}G(W_{_{TP}},m)-\frac{8}{15} W_{_{TP}}- \frac{8}{15}G(W_{_{TP}},m)\bigg( \frac{dG(W_{_{TP}},m)}{dW_{_{TP}}}\bigg)^{-1}=0\;,
\label{W G min}
\end{eqnarray}
and 
\begin{eqnarray}
\frac{\rho_{_{TP}}}{2}G(W_{_{TP}},m) + \frac{1}{3}G(W_{_{TP}},m)^2 - \frac{4}{15}W_{_{TP}} G(W_{_{TP}},m)=-\frac{3}{16}\rho_{_{TP}}^2\;.
\label{F_unwind W G}
\end{eqnarray}
where we have used $\gamma_{_U}=\sqrt{-\frac{3}{4}\rho_{_{TP}}}$ at the triple point. The above two equations can then be used to  solve for $\rho_{_{TP}}$ and $W_{_{TP}}$ in terms of $m$, yielding 
\begin{eqnarray}
\rho_{_{TP}}(m)=\frac{8}{19}\left[\pi m -4 - W_{_{TP}}(m)\ln(W_{_{TP}}(m))\right]\;,
\label{rhoTP}
\end{eqnarray}
where $W_{_{TP}}(m)$ is given by
\begin{eqnarray}
W_{_{TP}}(m)(\ln(W_{_{TP}}(m)))^2=\frac{19}{15}\left(4 -\pi m\right)\;.
\label{WTP}
\end{eqnarray}
Together these two equations give $\rho_{TP}$ as a function of $m$. We note that Eq.~(\ref{WTP}) implies that $m\leq m_1\equiv4/\pi$, and that both $W_{TP}(m)\rightarrow 0$ and $\rho_{TP}(m) \rightarrow 0$ as $m\rightarrow m_1$. For $m\lesssim m_1$ Eqs.~(\ref{rhoTP}) and (\ref{WTP}) can be approximated to give  $\rho_{_{TP}}$ explicitly in terms of $m$, i.e., Eq.~\ref{rhoTPm}.

\subsection{Locating the Phase Boundaries Near the Triple Point in Type I Systems}
\label{sec:LocatingPhaseBndyNearTriplePointTypeI}

Next we locate the first order phase boundaries that meet at the triple point. The boundary between the low ($L$) and high ($H$) tilt uniform states has already been found and lies along $h=0$ for $\rho<0$. The other two boundaries separate the modulated ($M$) and uniform high ($H$) tilt phases, and the modulated ($M$) and uniform low ($L$) tilt phases. Notationally we refer to the three phase boundaries as $L-H$, $M-H$, and $M-L$. The $M-H$ and $M-L$ phase boundaries are found by expanding near the triple point, i.e., for $\rho \lesssim \rho_{_{TP}}$. Specifically, expanding Eq.~(\ref{min W ito I})  for $W\approx W_{_{TP}}$, $h\approx h_{TP}=0$ and $\lvert\gamma\rvert \approx \lvert \gamma_{_{TP}} \rvert$ gives:
\begin{eqnarray}
m\pi=I(W_{_{TP}},h_{_{TP}}=0,\lvert \gamma_{_{TP}} \rvert)+I_W(m)\omega+I_\gamma(m)\nu+I_h(m)h\;,
\label{expanded I}
\end{eqnarray}
where $\omega=W-W_{_{TP}}$, $\nu=\lvert\gamma\rvert-\lvert \gamma_{_{TP}} \rvert$ and $\lvert \gamma_{_{TP}} \rvert$
\begin{eqnarray}
I_W(m)=\frac{dI}{dW}\bigg \rvert_{TP}\;, I_\gamma(m)=\frac{dI}{d\gamma_{_M}}\bigg \rvert_{TP}\;,  I_h(m)=\frac{dI}{dh}\bigg \rvert_{TP}\;.
\label{I derivs}
\end{eqnarray}
$I_W(m)$, $I_\gamma(m)$, and $I_h(m)$ depend on $m$ via $W_{_{TP}}(m)$ and $\lvert \gamma_{_{TP}} \rvert(m)$. We recall that $W_{_{TP}}(m)$ is given by Eq.~(\ref{WTP}) and $\lvert \gamma_{_{TP}} \rvert$ is obtained from Eq.~(\ref{gamma ito W})
\begin{eqnarray}
\lvert \gamma_{_{TP}} \rvert^2  = -\frac{3}{4}\rho_m+\frac{6}{19}\left[4\ln(2)+1-\ln(W_{_{TP}}) \right]W_{_{TP}}\;.
\label{gammaTP}
\end{eqnarray}
Since $m\pi=I(W_{_{TP}},h_{_{TP}}=0,\lvert \gamma_{_{TP}} \rvert)$, Eq.~(\ref{expanded I}) gives $\omega$ in terms of $\nu$ and $h$:
\begin{eqnarray}
\omega=-\frac{I_\gamma\nu+I_hh}{I_W}\;.
\label{omega}
\end{eqnarray}
Using this $\omega$ we insert $W=W_{_{TP}}+\omega$ into Eq.~(\ref{mod gamma abs min}) to obtain $\nu_\text{min}$:
\begin{eqnarray}
\nu_\text{min}=f_\rho(m) (\rho-\rho_{_{TP}})+f_h(m)h\;,
\label{nu_min}
\end{eqnarray}
where $f_\rho(m)$ and $f_h(m)$ are constants of order one. Inserting $\nu_\text{min}$ back into Eq.~(\ref{omega}) gives the corresponding $\omega_\text{min}$. The free energy near the triple point is then found by setting $W=W_{_{TP}}+\omega_\text{min}$ and $\lvert \gamma_{_{M}} \rvert=\lvert \gamma_{_{TP}} \rvert+\nu_\text{min}$ into Eq.~(\ref{F_unwind near critical point2}). To lowest order in $\rho-\rho_{_{TP}}$ and $h$ this gives
\begin{eqnarray}
\frac{F_M}{V}=\frac{-27u^3}{200 v^2} \bigg[ -\frac{3}{16}\rho_{_{TP}}^2+\frac{1}{2}\left[(\rho-\rho_{_{TP}})-h\right]\lvert \gamma_{_{TP}} \rvert^2\bigg].
\label{F_M near triple point}
\end{eqnarray}
We also minimize the uniform free energy of Eq.~(\ref{uniform energy}) for small $h$ and $\rho-\rho_{_{TP}}$:
\begin{eqnarray}
\frac{F_U}{V}=\frac{-27u^3}{200 v^2} \bigg[ -\frac{3}{16}\rho_{_{TP}}^2-\frac{3}{8}\rho_{_{TP}}(\rho-\rho_{_{TP}})
\mp\frac{\sqrt{-3\rho_{_{TP}}}}{2}h\bigg].
\label{F_U near triple point}
\end{eqnarray}
where the $\mp$ refers to $h>0$ (high tilt) and $h<0$ (low tilt) respectively. Setting $F_M=F_U$ gives the location of the first order phase boundaries. The modulated--high ($M-H$) tilt boundary and the modulated--low ($M-L$) tilt first order boundaries are:
\begin{eqnarray}
\rho_{_{M-H/M-L}}=\rho_{_{TP}}+\frac{\left(\lvert \gamma_{_{TP}} \rvert^2\mp\sqrt{-3\rho_{_{TP}}} \right)}{\left(\lvert \gamma_{_{TP}} \rvert^2+\frac{3}{4}\rho_{_{TP}} \right)}h\;.
\label{first order bndys}
\end{eqnarray}
We note that the magnitude of the boundaries' slopes are different, with the modulated--high tilt boundary being steeper. As $m$ is raised towards $m_1$, $W_{_{TP}}\ll 1$, and the boundaries can be approximated by Eq.~(\ref{approx first order bndys}).

\begin{acknowledgments}

JZ and KS acknowledge support from the National Science Foundation under Grant No. DMR-1005834.

\end{acknowledgments}


\begin{thebibliography}{99}
\bibitem{Sm-A* no macro chirality footnote} The Sm-$A$ phase, which is made up of the same chiral molecules, does not have a macroscopic chiral structure. Such a structure requires layering defects which replaces the Sm-$A$ phase with the twisted grain boundary (TGB) phase; see Ref.~\cite{RennLubensky}.
\bibitem{Non local U(1) footnote} The symmetry of the group of the superconducting-normal transition is actually local U(1) instead of the global U(1) symmetry of the prototypical $XY$ normal-superfluid transition.
\bibitem{polarization footnote} A more fundamental free energy would include a bilinear coupling between polarization and tilt, and also a bilinear coupling between polarization and field. Elimination (via energy minimization) of the polarization would result in the bilinear coupling between tilt and field that is presented here.
\bibitem{footnote on antiferroelectric or orthoconic Sm Cs} Examples of tilted phases with tightly wound helical modulations are antiferrolectric Sm-$C^*$ or orthoconic Sm-$C^*$ systems, in which the pitch length is just a few layers.
\bibitem{h not zero boundary} it can be shown using Eqs.~(\ref{F_u}) and (\ref{gamma min}) that the first order uniform low-- high tilt boundary deviates slightly from $h=0$ as $\rho$ is reduced.
\end{thebibliography}
\end{document}